\documentclass[%
 pyareprint,
nofootinbib,
 amsmath,amssymb,
 aps,
]{revtex4-2}
\usepackage{placeins}
\usepackage{graphicx}
\usepackage{dcolumn}
\usepackage{bm}
\usepackage{hyperref}
\usepackage{mathtools}\usepackage[utf8]{inputenc}
\usepackage{csquotes}
\usepackage{nccmath}
\usepackage{xcolor}
\usepackage{soul}
\usepackage{dirtytalk}

\usepackage{multirow}
\usepackage{lipsum}
\usepackage{makecell}
\usepackage{adjustbox}
\usepackage{amsmath}
\usepackage{resizegather}
\usepackage{rotating}
\begin{document}
\preprint{APS/123-QED}

 \title{Dynamical Wormhole Solutions in $f(R, T)$ Gravity}
\author{Yaghoub Heydarzade}
 \email{yheydarzade@bilkent.edu.tr}
\affiliation{
Department of Mathematics, Faculty of Sciences
Bilkent University, 06800 Ankara, Turkey.
}
\author{Maryam Ranjbar}
 \email{mranj006@ucr.edu}
\affiliation{ 
Department of Physics and Astronomy, University of California Riverside, 92521 California, USA.
}
\date{\today}

\begin{abstract}
A class of $f(R, T)$ theories extends the Einstein-Hilbert action by incorporating a general function of $R$ and $T$,  the Ricci scalar and the trace of the ordinary energy-momentum tensor $T_{\mu\nu}$, respectively, thereby introducing a specific modification to the Einstein’s field equations based on matter fields. Given that this modification is intrinsically tied to an energy-momentum tensor $T_{\mu\nu}$ that a priori respects energy conditions, we explore the potential of $f(R, T)$ theories admitting wormhole configurations satisfying energy conditions, unlike General Relativity, which typically necessitates exotic matter sources. Consequently, we investigate the existence of dynamical wormhole geometries that either uphold energy conditions or minimize their violations within the framework of trace of energy-momentum tensor squared gravity. To ensure the generality of our study, we consider two distinct equations of state for the matter content and systematically classify possible solutions based on constraints related to the wormhole's throat size, the coupling parameter of the theory, and the equation of state parameters.	
\end{abstract}

\maketitle

\section{\label{sec-Intro}Introduction\protect}

In spite of the accurate predictions of Einstein’s general relativity (GR) in many gravitational phenomena, it faces challenges in explaining the accelerated expansion of the universe and the existence of dark matter \cite{accelerateduni-1, accelerateduni-2}. One possibility for explaining the accelerated universe is the presence of an unknown form of energy with a repulsive effect, named as dark energy. Another possible explanation is that GR is insufficient on very large scales, and a more comprehensive theory is required to accurately describe the gravitational phenomena. In other words, by introducing an extension to GR, it may be possible to explain the accelerated expansion and structure formation of the universe without invoking dark energy or dark matter. Consequently, various attempts have been carried out to modify GR \cite{mgnutshell1, mgnutshell2}. Generally, the gravitational field equations depend on the nature of the matter source. Therefore, there is more than one way to generalize GR: it is possible to modify the geometrical part or the material part of GR’s field equations. For instance, Horndeski \cite{horndeski}, Einstein-Gauss-Bonnet \cite{GaussBonnet}, Einstein-Cartan theory \cite{ecartan}, and $f(R)$ theories \cite{intro2} are examples that originally generalize the geometrical side of Einstein’s field equations. On the other hand, Energy-Momentum Squared Gravity \cite{emsg} and quintessence models \cite{quintessence} are theories that generalize the material sector of Einstein’s field equations. In the context of cosmology, it has been shown that the field equations of any generic metric gravity theory in arbitrary dimensions are of the perfect fluid type \cite{Gurses2020,Gurses2024}. 
In $f(R)$ theories, first introduced in \cite{fr1, fr2, fr3},  the Einstein-Hilbert action is altered by including an arbitrary function of the Ricci scalar $R$.  Later, a generalized $f(R)$-type gravity model with arbitrary couplings in both geometry and matter parts was introduced in \cite{fr4}.
In \cite{fr5}, it was shown that in modified gravity theories with nonminimal coupling between matter and geometry, the specific form of the coupling dictates both the matter Lagrangian and the energy-momentum tensor. This results in an energy-momentum tensor that is more comprehensive than in standard GR, including an additional term dependent on the equation of state, which could be linked to internal stresses or other forms of internal energy. This insight led to the development of a theory with a maximal extension of the Hilbert-Einstein action in \cite{intro5}, named $f(R, \mathcal{L}_m)$, for wich $R$ is Ricci scalar and $\mathcal{L}_m$ is Lagrangian matter density. $f(R, \mathcal{L}_m)$ is able to modify both geometrical and material part of a gravitational theory at the same time.
A specific application of $f(R, \mathcal{L}_m)$ gravity is $\Lambda (T)$ gravity, which was introduced to describe an interacting dark energy. In this model, the cosmological constant $\Lambda$ in the gravitational Lagrangian is not a fixed value but a function of the trace of the stress-energy tensor $T$, aligning with recent cosmological data that support the idea behind $\Lambda (T)$ gravity \cite{frlambda}. Motivated by this concept, Harko et al. proposed $f(R,T)$ gravity in \cite{action}, where the gravitational Lagrangian incorporates a general function of the Ricci scalar $R$ and the trace of the stress-energy tensor $T$. Similar to $f(R, \mathcal{L}_m)$, $f(R,T)$ can simultaneously modify both the geometrical and material sector of Einstein’s field equations. 
Some interesting generalization of $f(R,T)$ theory are
$f(R, T, R_{\mu \nu} T^{\mu \nu})$ \cite{intro7} and $f(R, T^{\phi})$ \cite{intro8} with a self-interacting scalar field $\phi$.

The $f(R,T)$ gravity theory represents an extension beyond standard $f(R)$ gravity by introducing a non-minimal coupling between geometry and matter through the trace $T$ of the stress-energy tensor. The matter-geometry coupling introduces new phenomenology absent in both GR and $f(R)$ gravity. This $T$-dependence  may be induced by exotic imperfect fluids or quantum effects (conformal anomaly). As the result, it makes Newton's constant effectively matter-dependent, and generates non-geodesic motion by introducing an extra force \cite{action}. The theory maintains consistency by: (i) reducing to $f(R)$ gravity when $f_T=0$, (ii) preserving the Bianchi identities through modified conservation, and (iii) passing solar system tests \cite{action}.
The application of $f(R,T)$ gravity in addressing cosmological problems and understanding cosmic acceleration, as well as interactions between dark components, has been widely investigated in  \cite{intro9, intro10, intro11, intro12, intro13, intrort14, intrort15, intrort16, intrort17, intrort18, intrort19, intrort20, intrort21, intrort22}. Additionally, recent attention has also been given to stellar configurations and the astronomical aspects of this theory in order to study the consistency of the theory with observations \cite{intrort23, intrort24, intrort25, intrort26, intrort27, intrort28, intrort29, intrort30, intrort31}. Due to its aforementioned applications, we are interested in examining $f(R,T)$ theory versus the existence of dynamical wormhole solutions.

Wormholes demonstrate geometrical bridges between two separate regions of a universe or even two different universes. These structures could allow humans not only to travel between distant parts of a universes, or even two universes, but also to construct time machines. However, in GR, the flaring out condition on the throat of wormhole violates the weak energy condition (WEC). In other words, the existence of wormhole necessitates that energy density to be negative, and advent of a physically realistic wormhole requires large amount of exotic matter \cite{intro25, intro26,intro27}. Numerous research studies have made efforts to either fully resolve this problem or at least reduce the dependence on exotic matter to support the energy conditions \cite{intro27, intro28, intro29, intro30, intro31, intro32, intro33, intro35}.
One strategy involves creating thin-shell wormholes within the framework of GR through a cut-and-paste method. With this method, the exotic matter is concentrated at the throat; thereby reliance on exotic matter will be reduced \cite{intro35, intro36, intro37}. 
Another strategy involves exploring modified theories of gravity, where the modification terms in the given theory may allow for constructing wormhole structures without the presence of exotic matter and violation of energy conditions \cite{intro38, intro39}. In this respect, wormhole solutions in alternative theories such as $f(R)$ theory \cite{intro40, intro41, intro42, intro43, R2gravity}, scalar-tensor theory \cite{intro44}, energy-momentum squared gravity \cite{intro45, intro46-2}, and Rastall theory \cite{105, 106} have been studied.

The possibility of existence wormholes and corresponding energy conditions in $f(R, T)$ gravity has been studied in \cite{intrort71, intrort72, intrort73, intrort74}. Additionally, traversable wormhole solutions with different shape functions, coupling parameters, and equation of state parameters have been investigated in detail in \cite{intrort55, intrort56, intrort57, intrort58, intrort59, intrort60, intrort61, intrort62, intrort63, intrort64, intrort65, intrort66, intrort67}. These types of wormholes are of particular interest given that they allow a traveler to pass through the wormhole and exit safely. This requires the spacetime to be free of any event horizons and for the flaring-out condition to be satisfied. Within this framework, thin-shell wormholes also emerge as notable subjects of investigation \cite{intrort68, intrort69, intrort70}. However, the aforementioned research have mainly investigated static wormhole solutions. In contrast to static wormholes,  evolving wormholes provide a way to satisfy the WEC and create stable wormhole structures for a finite interval of time, for some pioneer works see \cite{intro29,intro30,intro47, intro49, intro50, intro53,intro55}. While the existence of static wormhole solutions in $f(R,T)$ gravity involving the trace-squared term $T^2$ has been studied in \cite{intrort57, intrort61, intro46}, we aim to investigate dynamical wormhole configurations supported by ordinary matter within the framework of the specific class $f(R,T)= R + \beta \, T^2$ where $\beta$ is the coupling parameter. Here, we develop a general approach that can be applied to all extended gravity theories which modify the matter sector of Einstein's field equations, building on the method introduced in \cite{intro59,105}. In the present study, the wormhole solution is described as dynamical due to its explicit dependence on time through the scale factor $a(t)$ in the spacetime metric. The motivation for introducing time dependence stems from several considerations:
\begin{enumerate}
\item  Cosmological Context: Embedding wormhole solutions within a cosmological background allows one to explore their evolution in time, potentially making them more realistic in the context of the expanding universe. This also helps to study how wormholes might behave or survive in an evolving spacetime, specially if they were formed in the early universe.   
\item Modified Gravity Framework: In 
$f(R,T)$ gravity, the coupling between matter and curvature naturally allows for non-conservation of the energy-momentum tensor. This extra interplay of the geometry and matter content leads to dynamic matter distributions and, consequently, time-dependent geometries. Thus, the structure of the theory itself supports and justifies the inclusion of dynamical solutions.
\item Choice of Stress-Energy tensor: If the matter source is modeled as a time-dependent perfect fluid or an anisotropic fluid with $\rho(t,r)$ and $p(t,r)$, the field equations naturally admit dynamical solutions. For instance, a time-varying pressure could stem from cosmological evolution or particle production effects in $f(R,T)$ theory.
\item Phenomenological Interest: Time-dependent wormholes are of considerable interest in the literature, as they may exhibit rich physical phenomena such as horizon formation/disappearance, energy condition violation/evolution, and interaction with cosmological fluids. 
\item Rahaman et al. recently proposed that galactic halos may exhibit properties conducive to supporting traversable wormholes \cite{rahaman}, prompting subsequent investigations into their potential detectability \cite{detect1, detect2, observingwh}. If such galactic wormholes do exist, it is reasonable to assume they were formed in the early Universe, thereby raising important questions regarding their physical characteristics and dynamical evolution throughout cosmic expansion. Even in the absence of observational evidence for these structures, the study of wormholes within a cosmological setting remains a topic of significant theoretical interest \cite{bahamonde}. Motivated by this, we explore the construction of evolving wormhole solutions in the context of $f(R, T)$ gravity under standard cosmological assumptions. The evolving wormhole metric is intentionally modeled to resemble the FLRW metric, ensuring that the wormhole evolves in tandem with the expansion of the Universe. The wormhole geometry is sustained by an anisotropic matter distribution that transitions to isotropy at large distances, in order to be compatible with the standard cosmological fluid description. Since isotropy about every point implies spatial homogeneity (though the converse is not necessarily true), the matter distribution becomes asymptotically homogeneous as well. This asymptotic regime may correspond to the cosmological horizon or, in the case of a sufficiently small wormhole, to a much smaller scale beyond which the spacetime approximates the FLRW geometry at a finite distance from the wormhole throat.
\end{enumerate}
While our present study primarily establishes the theoretical consistency of dynamical wormholes in $f(R,T)=R+\beta T^2$ gravity (e.g., WEC compliance, exact solutions) rather than observational tests, we would like to draw the interesting research in \cite{detect1, detect2, observingwh}, suggesting a framework for detecting traversable wormholes through astrophysical observations, to the eader's attention. For instance in \cite{observingwh}, the authors showed that if a wormhole connects two distinct spacetimes, then energy-momentum flux cannot be conserved independently within each region. Fields—scalar, electromagnetic, or gravitational—propagating near the wormhole in one spacetime are influenced by sources in the other. This non-local interaction is especially intriguing in the gravitational case: for instance, a star orbiting on the far side of a traversable wormhole connected to Sgr A* could affect the motion of the S2 star on our side, with potentially observable deviations detectable at upcoming precision levels. Similar signatures could arise in binary black hole or black hole–star systems. This also can be regarded as a key in distinction between wormholes and black holes - while both may contain similar mass concentrations, only wormholes would show anomalies dependent on dynamics occurring \enquote{on the other side}. This nonlocal influence closely parallels the time-dependent nature of the solutions presented in our work and the nonconservation of the energy-momentum tensor in
$f(R,T)$ gravity, where the coupling between matter and geometry leads to an exchange of energy-momentum between matter and the gravitational field. Just as a traversable wormhole enables interactions across spacetime domains, the non-conservation in $f(R,T)$ gravity reflects a breakdown of local conservation laws due to deeper geometric–matter coupling.

The organization of the present paper is as follows: in Section \ref{sec-fieldeq}, we introduce $f(R,T)$ gravity and its field equations. In Section \ref{sec-whgeometry}, we present dynamical Morris–Thorne wormhole spacetime and its governing field equations. Then, we provide exact solutions for the scale factor and shape function and perform a WEC analysis for two different equations of state (EoS) on the effective energy-momentum tensor in Sections \ref{generaleos} 
and \ref{particulareos}. In Section \ref{Conclusion}, our findings are concluded.
\section{\label{sec-fieldeq} \textbf{Trace of the energy-momentum tensor squared gravity} }
$f(R,T)$ gravity generalizes the Einstein-Hilbert action by replacing the Ricci scalar $R$ with a more general function $f(R,T)$ as follows
\begin{equation}
    \label{action1}
    \mathcal{S}=\frac{1}{2 \kappa}\int  \sqrt{-g} \, f(R,T)d^4 x+ \int  \sqrt{-g} \, L_m \,d^4 x,
\end{equation}
where $g$ is the determinant of the metric $g_{\mu \nu}$, $L_m$ is the matter Lagrangian, and $\kappa=8\pi G$. Here, $f(R,T)$ represents an arbitrary function of the Ricci scalar $R$, and trace $T$ of the ordinary energy-momentum tensor $T_{\mu \nu}$.

Following Ref. \cite{action}, the function $f(R, T)$ can be expressed in some different classes such as:
\begin{enumerate}
    \item $f(R,T)= R+ \alpha T,$
    \item $f(R,T)= f_{1}(R)+ f_{2}(T),$
    \item $f(R,T)= f_{1}(R)+ f_{2}(R) \, f_{3}(T),$
\end{enumerate}
where $\alpha$ is an arbitrary  constant and $f_1, f_2, f_3$ are arbitrary functions of their variables \cite{action}.  Key features of $f(R, T)$ theory can be noted as follows.
\begin{itemize}\item While GR and $f(R)$ gravity both feature minimal coupling between geometry and matter, the $f(R,T)$ theory generally introduces a non-minimal matter-geometry coupling directly at the level of the action. This non-minimal coupling modifies the field equations not only through higher-order curvature terms (as in $f(R)$) but also through direct matter content via $T$. Such a coupling can arise in the presence of exotic matter fields, effective quantum corrections, or semi-classical approximations where the trace anomaly plays a role. One may also consider linear forms such as cases 1 or 2 in the general action corresponding essentially to the Einstein-Hilbert action with an added scalar term, or to 
$f(R)$ gravity with a scalar contribution. However, it is important to emphasize that the nature of the introduced scalar plays a crucial role, as it can lead to markedly different physical implications. Different modified theories of gravity incorporate distinct scalar contributions, resulting in different equations of motion. For a formal discussion, see for instance the theorem regarding a generic gravitational action and the corresponding field equations established in \cite{GH} and \cite{GHS}. Moreover, the energy-momentum tensor in principle does not conserve in $f(R, T)$ gravity. This may be related to quantum effects \cite{action}, such as particle creation \cite{particle}, which is one of the motivations for including the $T$-dependence in the gravitational action. It is also worth mentioning that conserved versions of $f(R, T)$ theory have already been introduced \cite{non1, non2, non3}. Finally, it is worth to mention that in \cite{reexamine},  considering the case where the function \( f \) is separable, i.e., \( f(R, T) = f_1(R) + f_2(T) \), the  authors argue that the term \( f_2(T) \) should be interpreted as part of the matter Lagrangian \( \mathcal{L}_m \), and thus does not carry independent physical significance. Thereafter, a response to this critic was provided in \cite{hark} where the authors claim that they carefully reexamined the line of reasoning presented in the paper \cite{reexamine} and identified several significant conceptual issues concerning the author's physical interpretations, as well as the physical and mathematical methods employed in deriving the energy-momentum tensor of the theory.
\item At the level of equations of motion (EoM),  $f(R, T)$ gravity modifies the EoM of $f(R)$ gravity, as from Eqs. (11) in  \cite{action} one obtains
the field equations of $f(R)$ gravity when $f(R, T)\equiv f(R)$.  In particular, as mentioned before, the covariant divergence of the ordinary energy-momentum tensor $T_{\mu\nu}$ is non-zero in $f(R,T)$ gravity, see Eq.~(15) in \cite{action}. This leads to the presence of an extra force acting on test particles, and the motion is generally nongeodesic, Eqs.~(68) and (81)-(84) in  \cite{action}. Hence, $f(R, T)$ introduces novel matter-curvature interactions, leading to testable deviations from geodesic motion and modified conservation laws. The new matter and time dependent terms in the gravitational field equations induce a running gravitational coupling $G_{eff}$ and an effective cosmological constant $\Lambda_{eff}$, which can drive the late time acceleration of the FLRW universe. Also, in the limit $P \to 0$, corresponding to a pressureless fluid (dust), the motion of the test particles becomes geodesic akin to the standard GR. Interestingly, even if $f(R, T)\neq 0$, in this case the motion of the dust particles always follows the  geodesic lines of the FLRW geometry \cite{action}.
As we mentioned earlier, the role of $f(R,T)$ gravity in cosmological studies—particularly in addressing cosmic acceleration, dark sector interactions, and large-scale structure formation—has been extensively explored in \cite{intro9, intro10, intro11, intro12, intro13, intrort14, intrort15, intrort16, intrort17, intrort18, intrort19, intrort20, intrort21, intrort22}. More recently, attention has turned to its astrophysical implications, with research focusing on testing the theory’s consistency with observational data in \cite{intrort23, intrort24, intrort25, intrort26, intrort27, intrort28, intrort29, intrort30, intrort31}.
\end{itemize}

We shall consider a particular case of the above second class in this study. 
By applying the variational principle and assuming $L_m=\mathcal{P}$ in the action, 
the field equations of the second class can be obtained as \cite{action}
\begin{equation}
    \label{modifiedfieldeq}
    f_{1,R}(R) R_{\mu \nu}- \frac{1}{2}f_{1}(R) g_{\mu \nu}+(g_{\mu \nu} \Box -\nabla_{\mu} \nabla_{\nu})f_{1,R}(R)=\kappa T_{\mu \nu}-f_{2,T}(T) T_{\mu \nu}-f_{2,T}(T) (-2 T_{\mu \nu}+\mathcal{P} g_{\mu \nu})+\frac{1}{2}f_{2}(T) g_{\mu \nu},
\end{equation}
where $f_{1,R}(R)=\frac{df(R)}{dR}$ and $f_{2,T}(R)=\frac{df(T)}{dT}$. With the canonical choice of $f_{1}(R)=R$, the filed equations read as \cite{intro46}
\begin{equation}
    \label{fildeq1}
    G_{\mu \nu}= \kappa T_{\mu \nu}+ \frac{1}{2} f_2(T) g_{\mu \nu} + f_{2,T}(T) (T_{\mu \nu}- \mathcal{P} g_{\mu \nu}),
\end{equation}
where $G_{\mu \nu}$ and $T_{\mu \nu}$ represent the Einstein tensor and ordinary energy-momentum tensor, respectively. Assuming a generally anisotropic fluid, the ordinary energy-momentum tensor can be considered as
\begin{equation}
    \label{emtensor}
    T^{\mu}_{\nu}=diag\left(-\rho(t,r), P_r(t,r), P_l(t,r), P_l(t,r)\right).
\end{equation}
In this regard, the trace of energy-momentum tensor is $T=-\rho+ 3\mathcal{P}$, and the matter Lagrangian refers to the total pressure \cite{intro46}
\begin{equation}
    \label{totalp}
    \mathcal{P}=\frac{(P_r+ 2P_l)}{3}.
\end{equation}
In the context of trace of energy-momentum tensor squared gravity, we adopt a simple quadratic functional $f_{2}(T) = \beta \, T^2$, where $\beta$ is a constant coupling parameter, and later, we will show that it plays an important role in constructing wormholes with ordinary matter. For this choice of $f_{2}(T)$, the modified field equations \eqref{fildeq1} can be written as
\begin{equation}
    \label{fieldeq2}
    G_{\mu \nu}= \kappa T_{\mu \nu}^{eff},
\end{equation}
where
\begin{equation}
    \label{fieldeq3}
    \begin{split}
    T_{\mu \nu}^{eff}& = T_{\mu \nu}+\frac{1}{ \kappa} \left[ 2\beta (-\rho+ 3\mathcal{P}) \left(  T_{\mu \nu} +\frac{1}{4} (-\rho - \mathcal{P}) g_{\mu \nu}\right) \right].
    \end{split}
\end{equation}
It's important to highlight that here we have expressed the field equations of the trace of energy-momentum tensor squared theory using an effective energy-momentum tensor to facilitate checking the weak energy conditions. Interesting point for this consideration is that we will show although the effective energy-momentum tensor $T^{eff}_{\mu\nu}$ violates the WEC, the ordinary energy-momentum tensor $T_{\mu\nu}$ may respect it, that is in contrast to GR.  It is clear that any deviation from GR manifests on the right-hand side of the field equations \eqref{fieldeq2}, which describes the matter-energy content, rather than on the left-hand side, which encodes curvature. In this context, the additional terms in the effective energy-momentum tensor \eqref{fieldeq3} can be attributed to at least one of the following physical origins \cite{conservation2, conservation3, 104}: 
fluid imperfections (e.g. anisotropy), extra fluid, quantum effects (e.g. trace anomalies, renormalization effects), and an effective cosmological constant. The effective anisotropic energy-momentum tensor \eqref{fieldeq3} will possesses the following  effective density and pressures
\begin{equation}
    \label{5.1}
    \rho^{eff}=\rho+\frac{\beta  (2 P_l+P_r -\rho ) (2 P_l+P_r +15 \rho )}{6 \kappa },
\end{equation}
\begin{equation}
    \label{6.1}
    P_r^{eff} =\frac{r a(t)^2}{r-B(r)} \left(P_r+\frac{\beta  (2 P_l+P_r -\rho ) (12 P_r -(2 P_l+P_r +3 \rho ))}{6 \kappa } \right),
\end{equation}
\begin{equation}
    \label{7.1}
    P_l^{eff} =r^2 a(t)^2 \left(P_l+\frac{\beta  (2 P_l+P_r -\rho ) (12 P_l -(2 P_l+P_r +3 \rho ))}{6 \kappa } \right).
\end{equation} 
For the ordinary energy-momentum tensor components, we consider the following general linear EoS \cite{EOS1} 
\begin{equation}
    \label{50}
    2 P_l+P_r =\zeta \rho,
\end{equation}
where $\zeta\neq 1$ is a constant. From \eqref{fieldeq3}, we observe that for the particular case $\zeta =1$, the effect of $f(R, T)$ theory disappears and the theory reduces to GR.   Using the above EoS, the effective energy-momentum tensor components  \eqref{5.1}-\eqref{7.1} simplify to
\begin{equation}
    \label{51}
    \rho^{eff}=\rho\left(1+\rho \, \frac{\beta  (\zeta -1 ) (\zeta +15 )}{6 \kappa }\right),
\end{equation}
\begin{equation}
    \label{52}
        P_r^{eff} =\frac{r a(t)^2}{r-B(r)}\left(P_r+\frac{\rho \beta  (\zeta -1 ) (12 P_r - \rho \, (\zeta +3))}{6 \kappa }\right),
\end{equation}
    
\begin{equation}
    \label{53}
        P_l^{eff} =r^2 a(t)^2 \left(P_l+\frac{\rho \beta  (\zeta -1 ) (12 P_l-\rho \, (\zeta +3))}{6 \kappa } \right).
\end{equation}
\section{\label{sec-whgeometry} Dynamical wormhole
solution}
In this section, we elaborate on the field equations based on our metric, which describes a dynamical wormhole geometry. Specifically, we consider the time-dependent generalization of the Morris–Thorne metric as
\begin{equation}
    \label{5}
    ds^2 = -e^{2 \phi(r,t)} dt^2 +\mathit{a}(t)^2 \left( \frac{dr^2}{1-\frac{B(r)}{r}} + r^2 (d \theta^2 + sin^2 \theta \, d\psi^2) \right),
\end{equation}
where $a(t)$ is the scale factor, $e^{2 \phi(r,t)}$ denotes the redshift function, and $B(r)$ characterizes the shape function of the wormhole. Note that when $a(t)$ is a constant, this metric reduces to a static Morris-Thorne wormhole. Without any assumption on the form of $f(R, T)$ theory, using \eqref{fieldeq2}, we can derive the effective field equations for the aforementioned metric \eqref{5} and the general effective energy-momentum tensor $T^{\mu}_{\nu} \, ^{eff}=diag\left(-\rho^{eff}, P_r^{eff}, P_l^{eff}, P_l^{eff}\right)$ as follows

\begin{equation}
    \label{6}
    \kappa \rho^{eff}= 3 H^2 \, e^{-2 \phi (r,t)}+ \frac{B'(r)}{r^2 a(t)^2},
\end{equation}

\begin{equation}
    \label{7}
    \kappa P_r^{eff}= -e^{-2 \phi (r,t)} \left(3 H^2 + 2 \dot{H} -2 H \dot{\phi} \right) -\frac{B(r)}{r^3 a(t)^2}+ \frac{2 (r-B(r))}{r^2 a(t)^2} \phi',
\end{equation}

\begin{equation}
    \label{8}
    \kappa P_l^{eff} = -e^{-2 \phi (r,t)} \left(3 H^2 + 2 \dot{H} -2 H \dot{\phi} \right) +\frac{B(r)-r B'(r)}{2 r^3 a(t)^2} 
    +\frac{2 r-B(r) -r B'(r)}{2 r^2 a(t)^2} \phi' +\frac{(r-B(r))}{r a(t)^2} \left(\phi'^2+\phi'' \right),
\end{equation}

\begin{equation}
\label{9}
    0=2 H \phi',
\end{equation}
where $H=\frac{\dot{a}(t)}{a(t)}$, and the prime and dot signs represent derivation with respect to the radial coordinate $r$, and the temporal coordinate $t$, respectively. Given that the metric is dynamic (i.e., $\dot{a} (t) \ne 0$), equation \eqref{9} implies that $\phi' =0$, which means the redshift function is independent of the radial coordinate $r$. 
The following constraints on both the redshift and shape functions are necessary to ensure that the wormhole remains traversable and does not collapse or change into a different type of spacetime geometry: 
\begin{itemize}
\item [$\bullet$] There exists a minimum radial coordinate $r_0=B(r_0)$, defining the wormhole throat, connecting two asymptotic regions.\\
\item[$\bullet$] 
The shape function must respect the flaring-out condition $B(r)-r B'(r)>0$ near the throat, which simplifies to $B'(r_0)<1$ at the throat $r=r_0$.\\
\item[$\bullet$] For $r>r_0$, it is essential that the shape function satisfies the condition $1-\frac{B(r)}{r}>0$ to maintain the metric signature.\\
\item[$\bullet$] For asymptotically flat geometries,  
 the shape function $B(r)$ and redshift function $e^{2 \phi(t)}$ must hold the asymptotic flatness condition, i.e. $e^{2 \phi(t)} \to 1$, and $\frac{B(r)}{r} \to 0$ as $r \to \infty$.\\
\item[$\bullet$] To avoid the horizons and singularities, the redshift function $e^{2 \phi(t)}$ must be finite and nonzero everywhere.
\end{itemize}
Since $\phi' =0$, one can define a new time coordinate absorbing the $\phi(t)$ function. For the sake of simplicity, we shall keep the same notation for the new time coordinate resulting in $g_{00}=-1$. Thus, the absence of horizons throughout the spacetime is ensured in the present study.  However, black hole-wormhole transitions may occur for different metric constraints and gravitation theories. For instance, in \cite{7-1}, the authors examine static black hole-wormhole transitions in the context of braneworld gravity.

In addition, the Bianchi identity demands $\nabla_{\mu} {T^{eff~\mu}}_{\nu} \, =0$ for the effective energy-momentum tensor giving the following equations
\begin{equation}
    \label{10}
    \frac{\partial \rho^{eff}}{\partial t}+H (3 \rho^{eff}+ 2 P_l^{eff}+P_r^{eff} )=0,
\end{equation}

\begin{equation}
    \label{11}
    \frac{\partial P_r^{eff}}{\partial r}= \frac{2}{r} (P_l^{eff}-P_r^{eff} ).
\end{equation}
This implies breakdown of conservation of usual energy-momentum tensor $T_{\mu\nu}$  due to the non-minimal coupling in \eqref{fieldeq3}. Non-conservation of  $T_{\mu \nu}$ can be related to extra forces that changes the geodesic equation \cite{intro5, action}. From a cosmological perspective, this non-conservation could imply matter creation during the evolution of the universe, as discussed in\cite{conservation4}. The solutions later we obtain for the components of the ordinary energy-momentum tensor $T_{\mu \nu}$ follows Eqs. \eqref{10} and \eqref{11}. Regarding the metric considered, we aim to study evolving wormholes, with anisotropic and inhomogeneous sources, that smoothly merge with the cosmological background. This requires to integrate the present system of three nonlinear partial differential equations \eqref{6}-\eqref{8} with five unknowns: $a(t)$, $B(r)$, $\rho^{eff} (t,r)$, $P_r^{eff}(t,r)$, and $P_l^{eff}(t,r)$.  The matter source of these solutions will follow the dynamical equations \eqref{10} and \eqref{11}. Hence, fixing an effective equation of state, i.e., a physically viable constraint, is necessary for the derivation of general solutions.  One potential effective equation of state may involve only the set of unknown pressures $P_r^{eff}$ and $P_l^{eff}$, as investigated in \cite{intro59}. Alternatively, one can consider an effective equation of state in a more general form, involving $\rho^{eff}$, $P_r^{eff}$, and $P_l^{eff}$, as proposed in \cite{EOS1, EOS2, 105}.  For the sake of completeness of our present study, we investigate both these potential equations of state in detail in the following sections.

\section{\label{generaleos} Case I, EoS involving $\rho$, $P_r$, and $P_l$}
Here we consider a general equation of state that can encompass a wide range of physical systems. We adopt the following equation of state involving $\rho^{eff}$, $P_r^{eff}$, and $P_l^{eff}$, as proposed in \cite{EOS1, EOS2, 105},
\begin{equation}
    \label{12}
    \rho^{eff}(t,r)=\frac{\omega}{1+2 \gamma} \left(P_r^{eff} (t,r)+ 2 \gamma P_l^{eff}(t,r) \right),
\end{equation}
where $\omega$ and $\gamma$ are equation of state parameters. Depending on the values of $\omega$ and $\gamma$, this EoS can reduce to specific known equations of state as follows. It reduces to the EoS of traceless energy-momentum tensor as $-\rho^{eff}(t,r) +P_r^{eff}(t,r)+2P_l^{eff}(t,r)=0$ with $\omega=3$ and $\gamma=1$ \cite{104}. Additionally, it can represent the dimension $(d)$ dependent EoS $\rho^{eff}(t,r)=\alpha\left(P_r^{eff} (t,r)+ (d-2) P_l^{eff}(t,r) \right)$ for $d=4$ with $\gamma=1$ \cite{dim}, and the linear barotropic EoS $\rho^{eff}(t,r)=\omega P^{eff}(t,r)$ when $\omega=-1$ for all $\gamma$ with $P_r^{eff}(t,r)=P_l^{eff}(t,r)=P^{eff}(t,r)$. Hence, this EoS provides a framework for a more comprehensive understanding of a physical system describing a wormhole.

Substituting Eqs. \eqref{6},\eqref{7}, and \eqref{8} into \eqref{12}, applying the separation of variables, one obtains the following master equation
\begin{equation}
    \label{13}
    \begin{aligned}
    \frac{ -r B^{'}(r) \left( 1+ \gamma (2 + \omega) \right) \, - \omega (1- \gamma) B(r)}{  (1+2 \gamma)r^3} -\frac{(1+2 \gamma) \, a(t)^2\left( 2 \omega \dot{H} + 3 H^2 (\omega +1) \right)}{(1+ 2 \gamma) }=0.
    \end{aligned}
\end{equation}
This equation is separated into its radial and temporal parts, allowing each to be solved independently setting both terms to a constant $C$. This gives us the following ordinary differential equations
for $B(r)$ and $a(t)$ functions, respectively
\begin{equation}
    \label{14}
    \frac{ -r B^{'}(r) \left( 1+ \gamma (2 + \omega) \right) \, - \omega (1- \gamma) B(r)}{ (1+2 \gamma)r^3} = C,
\end{equation}
and
\begin{equation}
    \label{15}
    a(t)^2 \left[\left( - 2 \omega \right) \dot{H}+\left(- 3(\omega +1) \right)H^2 \right] =C.
\end{equation}
Utilizing these equations, one can find the shape function and scale factor for case $C=0$ and $C\ne0$ separately as follows.
\subsection{\label{sol-ce0-generaleos} General solution for the case $C=0$}
Upon integration of Eq. \eqref{14} with $C=0$, we obtain the shape function
\begin{equation}
    \label{29}
    B(r)= r_0 \left(\frac{r}{r_0}\right)^{\frac{(\gamma -1) \omega}{1+\gamma ( \omega+2)}},
\end{equation}
after applying the throat condition $B(r_0)=r_0$.
Similarly, integrating Eq. \eqref{15}  provides the scale factor
\begin{equation}
    \label{30}
    a(t)= (a_0 t+ a_1)^{\frac{2 \omega }{3 (\omega +1)}},
\end{equation}
where $a_0$ and $a_1$ represent integration constants. It can be seen that this solution evades the Big Bang singularity if $t\neq -\frac{a_1}{a_0}$.
\subsection{\label{sol-cne0-generaleos} General solution for the case $C\ne 0$}
Integrating Eq. \eqref{14} for $C\ne 0$ leaves us the following shape function
\begin{equation}
    \label{16.1}
        B(r)= \frac{C }{\omega+3}r^3+ C_1 \, r^{\frac{(\gamma -1) \omega}{1+\gamma ( \omega+2)}},
\end{equation}
where $C$ and $C_1$ are separation and integration constants, respectively. One can define the value of $C_1$ using the initial condition for the wormhole’s throat $B(r_0)=r_0$, as follows
\begin{equation}
    \label{17.1}
    C_1= \frac{\left(\omega+3- C  r_0^2\right)r_0 }{\left( \omega + 3\right)r_0^{\frac{(\gamma -1) \omega}{1+\gamma ( \omega+2)}}},
\end{equation}
from which, the flaring-out condition $B'(r_0)<1$ at the throat can be driven as
\begin{equation}
    \label{18.1}
    B'(r_0)=\frac{ C r_0^2(1+2 \gamma)  +\omega(\gamma- 1) }{1+\gamma ( \omega+2)}<1.
\end{equation}
Here, one notices that the value of the coefficient of the first term in \eqref{16.1}, i.e. $k=\frac{C }{-\omega-3}$, which depends on the set of parameters $\omega$ and $\gamma$, behaves like a cosmological constant. When $C$ is not equal to zero, the solutions behave similar to asymptotically  (anti) de Sitter spacetimes, rather than being asymptotically flat. This means that the asymptotic flatness condition is not satisfied when $C\ne 0$. 
Alternatively, we can write the $B(r)$ function as
\begin{equation}
        \label{19.1}
        B(r)= -k r^3 + B_n(r),
\end{equation}
where the parameter $k$ represents the spatial curvature of the background Friedmann-Lemaitre-Robertson-Walker (FLRW) metric, while $B_n(r)$ is the function describing the shape of a wormhole within this spacetime. The parameter $k$ takes values of $+1, -1$ or $0$, indicating closed, open, or flat universes, respectively.
It's important to distinguish between equations \eqref{16.1} and \eqref{19.1}, as discussed in references \cite{intro59, ext1, ext2}. In the latter, the throat condition $B_n(r_0)=r_0$ is applied only to the second term, $B_n(r)$, in the shape function, resulting in the constant $C_1$ calculated accordingly, i.e., we have $C_1=r_0^{1-\frac{(\gamma -1) \omega}{1+\gamma ( \omega+2)}}$ in this scenario. As mentioned in \cite{ext1, ext2}, applying the throat condition $B(r_0)=r_0$ places limitations on how far the wormhole solution can extend spatially, preventing it from being arbitrarily large
\begin{equation}
        \label{20}
        B'(r_0)=\frac{(\gamma -1) \omega}{1+\gamma ( \omega+2)}<1,
\end{equation}
while, the asymptotic flatness condition reads as 
\begin{equation}
        \label{21}
        -1<\frac{(1-\gamma ) \omega}{1+\gamma ( \omega+2)}<1.
\end{equation}
Moreover, in order to describe the geometry of the wormhole and its evolution over time, is necessary to obtain the scale factor, i.e., solution for $a(t)$ in Eq. \eqref{15} for $C\neq 0$.
\subsubsection{\label{generaleos-sol-R-cne0} Solution for the scale factor $a(t)$}

The equation \eqref{15} can be rewritten in a different but equivalent form as follows:
\begin{equation}
    \label{22}
    -2 \omega a (t) \ddot{a} (t)-(3 + \omega) \dot{a}^2= C.
\end{equation}
Integrating the above equation \eqref{22} yields the following first-order nonlinear differential equation
\begin{equation}
    \label{23}
    \dot{a}^2(t)=\frac{C}{-(3+ \omega)} \left( 1-a_0 \, a^{-\frac{3+\omega}{\omega}}\right),
\end{equation}
where $a_0$ is an integration constant. Consequently, the next integration gives
\begin{equation}
    \label{24}
    \int \frac{d a}{\sqrt{1-a_0 \, a^{-\frac{3+\omega}{\omega}}}}= \pm \sqrt{\frac{C}{-(3+ \omega)}}\int d t,
\end{equation}
where $\sqrt{\frac{C}{-(3+ \omega)}}>0$. Obtaining the explicit form of the scale factor for general $\omega$ is not possible. Hence, we derive its explicit solution for two specific cases: $\omega=-\frac{3}{2}$ and $\omega=-1$.
\begin{itemize}
    \item[$\bullet$] \label{omega1} \textit{ $\mathbf{\omega=-\frac{3}{2}}$}\\
    \\
In this case, Eqs. \eqref{24} and \eqref{16.1} give us the following form of the scale factor $a(t)$ and shape function $B(r)$ 
    \begin{equation}
        \label{25}
        \begin{aligned}
            a(t)& =\frac{1}{a_0}-\frac{a_0}{4}\left(\pm \sqrt{\frac{-2 C}{3}}t + a_1\right)^2, \\ 
            B(r) & =\frac{2}{3} C r^3 -\frac{r_0 \left(3-2 C r_0^2\right)}{3} \, \left(\frac{r}{r_0}\right)^{\frac{-3 (\gamma -1)}{\gamma +2}},
        \end{aligned}
    \end{equation}
where $a_1$ is an integration constant.\\   
Alternatively, by considering Eq. \eqref{19.1} as the shape function, the $B_n(r)$ elucidates form of the shape function for the inhabiting wormhole. Therefore, the scale factor and shape function read as
    \begin{equation}
        \label{26}
        \begin{aligned}
            &a(t) =\frac{1}{a_0}-\frac{a_0}{4}\left(\pm \sqrt{k}t + a_1\right)^2, \\
            &B_n(r) = r_0 \, \left(\frac{r}{r_0}\right)^{\frac{-3 (\gamma -1)}{\gamma +2}},
        \end{aligned}
    \end{equation}
    where $k=\frac{C }{-3/2}$, and ensuring the validity of the solution necessitates $k=+1$ , and thus $C=-\frac{3}{2}$. Hence, in this case, the existence of a wormhole solution requires a positively curved spatial hypersurface.
    \\
    \item[$\bullet$] \label{omega2} 
    \textit{$\mathbf{\omega=-1}$}\\
    \\
Utilizing Eqs. \eqref{16.1} and \eqref{24}, we can characterize the behavior of the scale factor and shape function when $\omega=-1$
    \begin{equation}
        \label{27}
        \begin{aligned}
            & a(t)= \frac{1}{\sqrt{a_0}}\sin \left(\pm \sqrt{\frac{Ca_0}{-2}}\,t+a_1\right),\\
            & B(r)= -\frac{1}{2}C r^3 - \frac{r_0 \left(-2 + C r_0^2 \right)}{2} \, \left(\frac{r}{r_0}\right)^{\frac{1-\gamma}{1+\gamma}}.
        \end{aligned}
    \end{equation}
    \\
Note that $R_{1}$ is an integration constant. On the other hand, adopting the Eq. \eqref{19.1} as the shape function provides following solutions for the scale factor and shape function 
    \begin{equation}
        \label{28}
        \begin{aligned}
            & a(t)= \frac{1}{\sqrt{a_0}}\sin \left(\pm \sqrt{k a_0}\,t+a_1\right),\\
            & B_{n}(r)= r_0 \, \left(\frac{r}{r_0}\right)^{\frac{1-\gamma}{1+\gamma}}.
        \end{aligned}
    \end{equation}
\end{itemize}
\subsection{\label{wecc=0}Analyzing the Weak Energy Condition}
In this section, we investigate whether the previously obtained solutions are supported by ordinary matter sources respecting the WEC.  
Equating Eqs. \eqref{6}, \eqref{7}, and \eqref{8} with $\rho^{eff}$,$P_r^{eff}$, $P_l^{eff}$ from Eqs. \eqref{51}, \eqref{52}, and \eqref{53}, respectively, one finds the ordinary energy-momentum tensor components as follows
\begin{equation}
    \label{rho3.41}
    \rho(t,r)=\frac{\kappa  \left(\sqrt{\frac{6 \beta  (\zeta -1) (\zeta +15) \left(B'(r)+3 r^2 \dot{a}(t)^2\right)}{\kappa ^2 r^2 a(t)^2}+9}-3\right)}{\beta  (\zeta -1) (\zeta +15)},
\end{equation}

\begin{equation}
    \label{pr3.42}
    P_r(t,r)=\frac{6 r B(r) \left(r^2 \left(2 a(t) \ddot{a}(t)+\dot{a}(t)^2\right)-1\right)+6 B(r)^2+r^4 \left(-12 a(t) \ddot{a}(t)-6 \dot{a}(t)^2+\beta  (\zeta -1) (\zeta +3) \rho(t,r)^2 a(t)^4\right)}{6 r^4 a(t)^4 (2 \beta  (\zeta -1) \rho(t,r) +\kappa )},
\end{equation}

\begin{equation}
    \label{pl3.43}
    P_l(t,r)=\frac{-3 r \left(B'(r)+2 r^2 \dot{a}(t)^2\right)+3 B(r)+\beta  (\zeta -1) (\zeta +3) \rho(t,r)^2 r^5 a(t)^4-12 r^3 a(t) \ddot{a}(t)}{6 r^5 a(t)^4 (2 \beta  (\zeta -1) \rho(t,r) +\kappa )}.
\end{equation}
In the following, we analyze the flaring-out and WE conditions for two different cases $C=0$ and $C\neq 0$.
\begin{itemize}
    \item[$\bullet$] \label{c0casee} 
\textit{$\mathbf{C=0}$}\\
    \\
To examine the case $C=0$, substituting the scale factor \eqref{29} and shape function \eqref{30} into above equations, one finds following relations
\begin{equation}
    \label{rhocas1}
    \rho=\frac{\kappa  \left(\sqrt{\chi_1+9}-3\right)}{\beta  (\zeta -1) (\zeta +15)},
\end{equation}

\begin{equation}
    \label{prcase1}
    \begin{aligned}
        \rho+P_r=&\frac{\kappa  \left(\sqrt{\chi_1+9}-3\right)}{\beta  (\zeta -1) (\zeta +15)}+ \frac{(\zeta +15) (a_0 t+a_1)^{-\frac{8 \omega }{3 (\omega +1)}}}{6 \kappa  r^4 \left(\zeta +2 \sqrt{\chi_1+9}+9\right)}
        \left(8 r^3 a_0^2 \omega  \left(r-r_0 \left(\frac{r}{r_0}\right)^{\frac{(\gamma -1) \omega }{\gamma  (\omega +2)+1}}\right) (a_0 t+a_1)^{-\frac{2 (\omega +3)}{3 (\omega +1)}} \, (\omega +1)^{-2} \right.\\
        &\left.-\frac{(\zeta +3) \kappa ^2 r^4 \left(6 \left(\sqrt{\chi_1+9}-3\right)-\chi_1 \right) (a_0 t+a_1)^{\frac{8 \omega }{3 (\omega +1)}}}{\beta  (\zeta -1) (\zeta +15)^2}+6 r_0 \left(\frac{r}{r_0}\right)^{\frac{(\gamma -1) \omega }{\gamma  (\omega +2)+1}} \left(r_0 \left(\frac{r}{r_0}\right)^{\frac{(\gamma -1) \omega }{\gamma  (\omega +2)+1}}-r\right)\right),
    \end{aligned}
\end{equation}
\begin{equation}
    \label{plcas1}
    \begin{aligned}
    \rho+P_l=\frac{\kappa  \left(\sqrt{\chi_1+9}-3\right)}{\beta  (\zeta -1) (\zeta +15)}+\frac{(\zeta +15) (a_0 t+a_1)^{-\frac{8 \omega }{3 (\omega +1)}}}{6 \kappa  r^5 \left(\zeta +2 \sqrt{\chi_1+9}+9\right)}\Bigg( &\frac{(\zeta +3) \kappa ^2 r^5 \left(6 \left(\sqrt{\chi_1+9}-3\right)-\chi_1 \right) (a_0 t+a_1)^{\frac{8 \omega }{3 (\omega +1)}}}{(\zeta +15)^2 (\beta -\beta  \zeta )} \\
    &  +\frac{8 r^3 a_0^2 \omega  (a_0 t+a_1)^{-\frac{2 (\omega +3)}{3 (\omega +1)}}}{(\omega +1)^2}+\frac{3 r_0 (2 \gamma +\omega +1)}{\gamma  (\omega +2)+1} \left(\frac{r}{r_0}\right)^{\frac{(\gamma -1) \omega }{\gamma  (\omega +2)+1}} \Bigg),
    \end{aligned}
\end{equation}
where $\chi_1$ is defined as
\begin{equation}
    \label{chicas1}
    \chi_1=\frac{6 \beta  (\gamma -1) (\zeta -1) (\zeta +15) r_0 \omega  \left(\frac{r}{r_0}\right)^{\frac{(\gamma -1) \omega }{\gamma  (\omega +2)+1}} (a_0 t+a_1)^{-\frac{4 \omega }{3 (\omega +1)}}}{\kappa ^2 r^3 (\gamma  (\omega +2)+1)}+\frac{8 \beta  (\zeta -1) (\zeta +15) a_0^2 \omega ^2}{\kappa ^2 (\omega +1)^2 (a_0 t+a_1)^2}.
\end{equation}
In this case, the flaring-out, flatness, and weak energy
condition ($\rho \ge 0$, $\rho+P_r \ge 0$, and $\rho+P_l \ge 0$) can all be satisfied simultaneously for both of the following possibilities
\begin{fleqn}
\begin{equation}
    \label{weccas1c01}
    \begin{split}
       &\beta <0, ~~ -15<\zeta <1, \,
    \end{split}
    \begin{cases}
    a_0>0, \,a_1>0, ~~ \omega<-3, ~~ \frac{\omega -1}{2 \omega +2}<\gamma <\frac{1}{2} (-\omega -1); &~~~~ r_0 > \frac{\sqrt{3}}{2} \sqrt{\frac{(\gamma -1) a_1^2 (\omega +1)^2 a_1^{-\frac{4 \omega }{3 (\omega +1)}}}{a_0^2 (\gamma  (\omega +2) \omega +\omega )}}, \\
    a_0>0, \,a_1>0, ~~ -3<\omega<-2,~~ \frac{1}{2} (-\omega -1)<\gamma <\frac{\omega -1}{2 \omega +2}; &~~~~ r_0>\frac{\sqrt{3}}{2} | \omega +1|  \sqrt{\frac{(\gamma -1) a_1^2 a_1^{-\frac{4 \omega }{3 (\omega +1)}}}{a_0^2 (\gamma  (\omega +2) \omega +\omega )}},  \\
    a_0>0, \,a_1>0, ~~ -2 \leq \omega<-1,~~ \frac{1}{2} (-\omega -1)<\gamma <\frac{\omega -1}{2 \omega +2} ; &~~~~ r_0 \geq \frac{\sqrt{6}}{4} \sqrt{-\frac{(\omega +1)^2 (2 \gamma +\omega +1) a_1^{-\frac{4 \omega }{3 (\omega +1)}}}{a_0^2 \omega  (\gamma  (\omega +2)+1) a_1^{-\frac{2 (5 \omega +3)}{3 (\omega +1)}}}},
    \\
    a_0>0, \,a_1>0, ~~ \omega>0, ~~~ \frac{\omega -1}{2 \omega +2}<\gamma \leq 1; & ~~~~~ r_0>\frac{1}{2} \sqrt{3} \sqrt{\frac{(\gamma -1) a_1^2 (\omega +1)^2 a_1^{-\frac{4 \omega }{3 (\omega +1)}}}{a_0^2 (\gamma  (\omega +2) \omega +\omega )}},
    \end{cases}
    \end{equation}
\end{fleqn}
or
\begin{fleqn}
\begin{equation}
    \label{weccas1c02}
    \begin{split}
       &\beta >0, ~~\zeta >1,\,
    \end{split}
    \begin{cases}
    a_0>0, \,a_1>0, ~~ \omega<-3, ~~ \frac{\omega -1}{2 \omega +2}<\gamma <\frac{1}{2} (-\omega -1); &~~~~ r_0 > \frac{\sqrt{3}}{2} \sqrt{\frac{(\gamma -1) a_1^2 (\omega +1)^2 a_1^{-\frac{4 \omega }{3 (\omega +1)}}}{a_0^2 (\gamma  (\omega +2) \omega +\omega )}}, \\
    a_0>0, \,a_1>0, ~~ -3<\omega<-2,~~ \frac{1}{2} (-\omega -1)<\gamma <\frac{\omega -1}{2 \omega +2}; &~~~~ r_0>\frac{\sqrt{3}}{2} | \omega +1|  \sqrt{\frac{(\gamma -1) a_1^2 a_1^{-\frac{4 \omega }{3 (\omega +1)}}}{a_0^2 (\gamma  (\omega +2) \omega +\omega )}},  \\
    a_0>0, \,a_1>0, ~~ -2 \leq \omega<-1,~~ \frac{1}{2} (-\omega -1)<\gamma <\frac{\omega -1}{2 \omega +2} ; &~~~~ r_0 \geq \frac{\sqrt{6}}{4} \sqrt{-\frac{(\omega +1)^2 (2 \gamma +\omega +1) a_1^{-\frac{4 \omega }{3 (\omega +1)}}}{a_0^2 \omega  (\gamma  (\omega +2)+1) a_1^{-\frac{2 (5 \omega +3)}{3 (\omega +1)}}}},
    \\
    a_0>0, \,a_1>0, ~~ \omega>0, ~~~ \frac{\omega -1}{2 \omega +2}<\gamma \leq 1; & ~~~~~ r_0>\frac{1}{2} \sqrt{3} \sqrt{\frac{(\gamma -1) a_1^2 (\omega +1)^2 a_1^{-\frac{4 \omega }{3 (\omega +1)}}}{a_0^2 (\gamma  (\omega +2) \omega +\omega )}}.
    \end{cases}
    \end{equation}
\end{fleqn}\\
\\ 

Also, considering the scale factor in \eqref{30}, i.e $a(t) = (a_0 t + a_1)^{\frac{2\omega}{3(\omega + 1)}}$, where $a_0$, $a_1$, and $\omega \neq -1$ are constants, to remain real and positive, condition $a_0 t + a_1 > 0$ must hold, implying $t > -\frac{a_1}{a_0}$ if $a_0 > 0$, or $t < -\frac{a_1}{a_0}$ if $a_0 < 0$. The sign of $\dot a(t)=\frac{2\omega}{3(\omega + 1)} a_0 (a_0 t + a_1)^{\frac{2\omega}{3(\omega + 1)} - 1}$ determines whether the wormhole is expanding or contracting. The sign of $\dot{a}(t)$ depends on both $a_0$ and $\omega$:
\begin{itemize}
    \item If $\frac{2\omega}{3(\omega + 1)} > 0$ and $a_0 > 0$, then $\dot{a}(t) > 0$: expanding wormhole.
    \item If $\frac{2\omega}{3(\omega + 1)} < 0$ and $a_0 > 0$, then $\dot{a}(t) < 0$: contracting wormhole.
\end{itemize}
Hence, if $a_0>0$, for typical matter sources such as dust ($\omega = 0$) or radiation ($\omega = 1/3$), the wormhole expands as a power law in time. If $\omega < -1$, corresponding to phantom energy, the scale factor exhibits super-accelerated expansion. Also, when $ -1<\omega<0$, the WEC is not satisfied for the solutions obtained, and hence this range should be excluded. At the critical time $t = -\frac{a_1}{a_0}$, the scale factor either vanishes or diverges, indicating a cosmological singularity.

 The behavior of $\rho$, $\rho+P_r$, and $\rho+P_l$, along with $B(r)/r$, is illustrated in Figure \ref{fig1-brr} for a specific set of parameters satisfying the above constraints. Specifically, we consider the case where the EoS \eqref{12} reduces to the EoS of the traceless energy-momentum tensor, corresponding to 
$\omega=3$ and $\gamma=1$. Notably, $\omega=3$ represents ordinary matter, and Figure \ref{fig1-brr} demonstrates that within the framework of
$f(R,T)$ gravity, a wormhole can be supported by ordinary matter, without requiring exotic matter. This is in contrast to the behavior of the effective energy-momentum tensor which violates the WEC. In the late-time limit  ($t \to \infty$), the exact solutions we obtained indicate that the energy density $ \rho$ and the pressure components $P_r $ and $ P_l $ asymptotically vanish for $a_0>0$. This behavior is physically consistent, as the corresponding exact solution for the scale factor exhibits growth with time, reflecting an expanding wormhole spacetime for $a_0>0$. The behavior of $\rho$, $P_r$, and $P_l$ as $t \to \infty$ is shown in Fig \ref{re-fig1infty}.\\
\begin{figure*}[htb!]
\begin{minipage}{\textwidth}
\includegraphics[width=\textwidth]{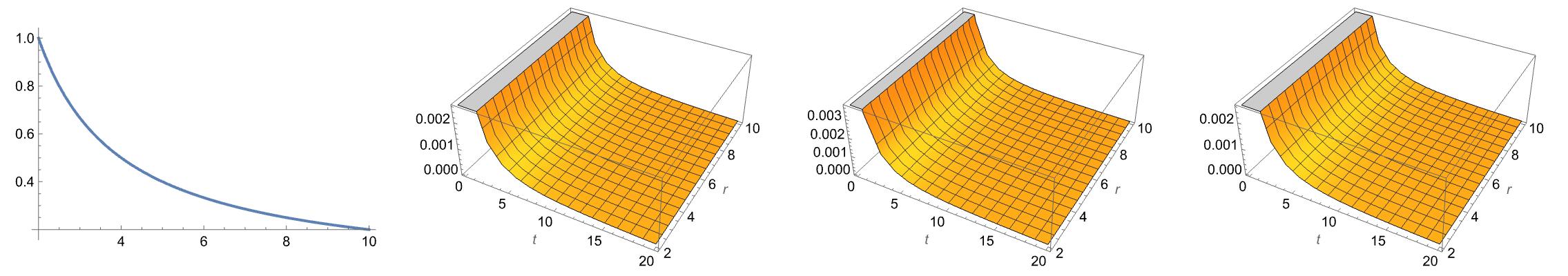}
\caption{\label{fig1-brr} This figure indicates the behavior of $B(r)/r$, $\rho$, $\rho+P_r$ , and $\rho+P_l$ regarding r and t from left to right, respectively,for $2 < r < 10$ and $0 < t < 20$, considering Eqs.\eqref{rhocas1}-\eqref{chicas1}.
According to \eqref{weccas1c02}, the arbitrary constants are taken as follows:$\beta=2$, $\zeta=2$, $\gamma=1$, $\omega=3$, $a_0=3$, $a_1=3$, $\kappa=8 \pi$, and $r_0=2$.}
\end{minipage}
\end{figure*}
\begin{figure*}[htb!]
\begin{minipage}{\textwidth}
\includegraphics[width=0.75\textwidth]{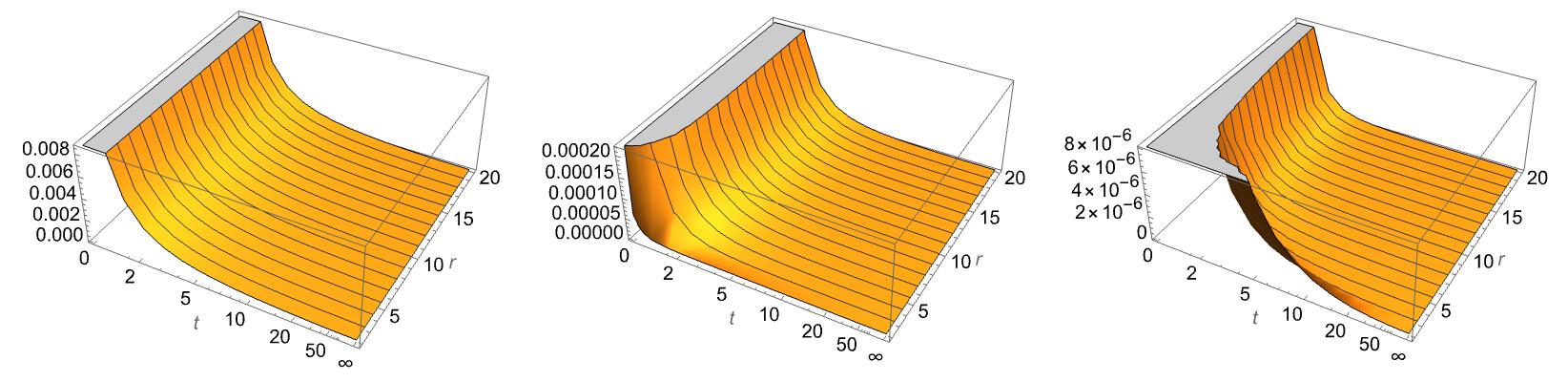}
\caption{\label{re-fig1infty} This figure indicates the behavior of $\rho$, $P_r$ , and $P_l$ with respect to r and t from left to right, respectively,for $2 < r < 10$ and $0 < t < \infty$, corresponding to Eqs.\eqref{rho3.41}-\eqref{pl3.43}, while $C=0$ and the shape function and scale factor are considered as \eqref{29} and \eqref{30} respectively.
According to \eqref{weccas1c02}, the arbitrary constants are taken to be the same as in Fig \ref{fig1-brr}, which are:$\beta=2$, $\zeta=2$, $\gamma=1$, $\omega=3$, $a_0=3$, $a_1=3$, $\kappa=8 \pi$, and $r_0=2$.}
\end{minipage}
\end{figure*}
\newpage
\item[$\bullet$] \label{cne0casee} 
\textit{$\mathbf{C\ne0}$}\\
\\
    In our previous work \cite{105}, we showed that the  wormhole solutions with $C\neq 0$, outlined in section \ref{sol-cne0-generaleos}, satisfies the WEC in Rastall theory. In contrast, when the same solution is examined within the framework of $f(R,T)$ theory, it does not meet the WEC. This discrepancy not only underscores the physical viability of our solution in capturing the essential features of wormhole geometries in distinct modified gravity models but also highlights the sensitivity of energy condition constraints to the specific modifications imposed on the source terms of the field equations. Hence, here we only investigate WEC for the general solution with $C=0$.
\end{itemize}
\section{\label{particulareos} Case II, EoS involving $P_r$ and $P_l$}
Here, we adopt the alternative equation of state used in \cite{intro59} for the effective energy-momentum tensor as follows
\begin{equation}
    \label{31}
    P_r^{eff}(r, t) = \alpha \, P_l^{eff}(r, t),
\end{equation}
where $\alpha$ can be either a constant or gnerally a function of $r$.
With the same methodology as in the previous section, by inserting Eqs. \eqref{7} and \eqref{8} into the above EoS \eqref{31} and then separating the temporal and radial components, we derive the following master equations
\begin{equation}
    \label{33}
    a(t)^2 \left(3 H^2+2 \dot{H}\right)=C,
\end{equation}
\begin{equation}
    \label{34}
    B'(r)-\frac{B(r)}{r}\left( 1+ \frac{2}{\alpha(r)}\right)-2 C \, r^2 \left(\frac{1}{\alpha(r)}-1 \right)=0.
\end{equation}
One can immediately find the scale factor function by integrating Eq. \eqref{33} for $C=0$ as
\begin{equation}
    \label{41}
    a(t)=(a_0 \, t - a_1)^{2/3},
\end{equation}
where $a_0$ and $a_1$ are integration constants. In light of the lack of an analytical solution for Eq. \eqref{33} when $C\neq 0$, here, we will restrict our analysis for WEC only to the case $C=0$.
\subsection{\label{solB-particulareos} General solution for the shape function}
Regardless of the specific form of the function $\alpha(r)$, we can always represent it using another arbitrary function $s(r)$ in a particular way, introduced in \cite{intro59}
\begin{equation}
    \label{alphagr}
    \alpha(r)= \frac{s(r)}{r s'(r)}.
\end{equation}
Using \eqref{alphagr}, we can obtain the shape function $B(r)$ in terms of $s(r)$ by integrating Eq. \eqref{34} as
\begin{equation}
    \label{35}
    B(r)=-C r^3+r \, s(r)^2.
\end{equation}
Here, the constant $-C$ characterizes the spatial curvature of spacetime, representing potential values of $0$, $-1$, or $+1$ which respectively correspond to a flat, open, or closed FLRW solutions. We also observe that as the function $s(r)^2$ approaches zero as $r$ tends towards infinity, the resulting metric resembles the FLRW spacetime. For the case $C=0$, the shape function reduces to 
\begin{equation}
    \label{bng}
    B_n(r)=r s(r)^2.
\end{equation}
This expression represents the effective shape function, which characterizes the wormhole throat as described in \cite{intro59}

We consider the following forms for $s(r)$ proposed in \cite{intro59} and obtain the corresponding shape functions as follows.

\begin{itemize}
\item[$\bullet$] \label{g1} \textit{ $\mathbf{s(r)=\left(\frac{r}{r_0} \right)^{1/\alpha}}$}\\
        \\

By setting $\alpha=constant$, the effective shape function \eqref{bng} reduces to
\begin{equation}
    \label{37}
    B_n(r)=r \left(\frac{r}{r_0} \right)^{2/\alpha},
\end{equation}
where $r_0$ is integration constant and flaring-out condition holds for $\alpha<0$.
\item[$\bullet$] \label{g2} \textit{$\mathbf{s(r)=\left(\frac{r}{r_0} \right)^{n/2}}$}\\
        \\
 In this case, we have the effective shape function as follows
\begin{equation}
    \label{38}
    B_n(r)=r \left(\frac{r}{r_0} \right)^{n},
\end{equation}
where $r_0$ is constant and the flaring-out condition is respected when $n<0$.
\\
    \item[$\bullet$] \label{g3}\textit{$\mathbf{s(r)=\sqrt{\frac{1}{1+r-r_0}}}$}\\
\\
In the case of a more general $\alpha$ being a function of $r$, we can consider the scenario where $s(r)$ assumes this form, resulting in following effective shape function
\begin{equation}
    \label{39}
    B_n(r)=r \left( \frac{1}{1+r-r_0} \right).
\end{equation}
Here, the flaring-out condition will be satisfied for all $r_0>0$.
\\
        \item[$\bullet$] \label{g4} 
        \textit{$\mathbf{s(r)=\sqrt{\epsilon (1-\frac{r_0}{r})+\frac{r_0}{r}}}$}\\
        \\
In this scenario, the corresponding effective shape function is determined as follows
\begin{equation}
    \label{40}
    B_n(r)=r \left( \epsilon \, (1-\frac{r_0}{r})+\frac{r_0}{r} \right),
\end{equation}
where $\epsilon \leq 1$ ensures the satisfaction of the flaring-out condition at the throat  $r=r_0$.

A straightforward verification demonstrates that all the aforementioned examples satisfy the condition $B_n(r_0) = r_0$, thereby confirming the presence of a viable wormhole throat at  $r_0$.

\end{itemize}

\subsection{\label{EM-particulareos} Analyzing the Weak Energy Condition}
Employing the effective EoS \eqref{31} coupled with the conservation law Eq.\eqref{11} allows us to derive solutions for $P_r^{eff}$ and $P_l^{eff}$, while one cand find $\rho^{eff}$ from \eqref{6} as follows
\begin{equation}
    \label{46}
    \rho^{eff}(t,r)=\frac{1}{\kappa a(t)^2} \left(3 \dot{a}(t)^2 + \frac{B'(r)}{r^2} \right),
\end{equation}
\begin{equation}
    \label{48}
    P_r^{eff}(t,r)= P^{*}(t)  \frac{s(r)^2}{r^2},
\end{equation}
\begin{equation}
    \label{49}
    P_l^{eff}(t,r)= P^{*}(t) \frac{ s(r) s'(r)}{r}.
\end{equation}
Using equations \eqref{7} and \eqref{8} for $P_r^{eff}$ and $P_l^{eff}$, alongside equations \eqref{33} and \eqref{34}, one can find
\begin{equation}
    \label{P11}
    P^{*}(t)=-\frac{1}{\kappa \,  a(t)^2}.
\end{equation}
To obtain the ordinary energy-momentum tensor components within this framework of $f(R, T)$ theory under the EoS \eqref{31}, we substitute $\rho^{eff}$ from \eqref{46}, $P_r^{eff}$ from \eqref{48}, and $P_l^{eff}$ from \eqref{49} into the system of equations \eqref{51}, \eqref{52}, and \eqref{53}. Subsequently, by incorporating the shape function \eqref{bng} we find
\begin{equation}
    \label{rhocase265}
    \rho(t,r)=\frac{\kappa  \left(\sqrt{\frac{6 \beta  (\zeta -1) (\zeta +15) \left(3 r^2 \dot{a}(t)^2+2 r s(r) s'(r)+s(r)^2\right)}{\kappa ^2 r^2 a(t)^2}+9}-3\right)}{\beta  (\zeta -1) (\zeta +15)},
\end{equation}
\\
\begin{equation}
    \label{prcase266}
    P_r(t,r)=\frac{(\zeta +15) \left(\frac{(\zeta +3) \kappa ^2 r^3 a(t)^4 \left(\sqrt{\frac{6 \beta  (\zeta -1) (\zeta +15) \left(3 r^2 \dot{a}(t)^2+2 r s(r) s'(r)+s(r)^2\right)}{\kappa ^2 r^2 a(t)^2}+9}-3\right)^2}{\beta  (\zeta -1) (\zeta +15)^2}+6 r s(r)^2 \left(s(r)^2-1\right)\right)}{6 \kappa  r^3 a(t)^4 \left(\zeta +2 \sqrt{\frac{6 \beta  (\zeta -1) (\zeta +15) \left(3 r^2 \dot{a}(t)^2+2 r s(r) s'(r)+s(r)^2\right)}{\kappa ^2 r^2 a(t)^2}+9}+9\right)},
\end{equation}
\\
\begin{equation}
    \label{Plcase267}
    P_l(t,r)=\frac{(\zeta +15) \left(\frac{(\zeta +3) \kappa ^2 r^3 a(t)^4 \left(\sqrt{\frac{6 \beta  (\zeta -1) (\zeta +15) \left(3 r^2 \dot{a}(t)^2+2 r s(r) s'(r)+s(r)^2\right)}{\kappa ^2 r^2 a(t)^2}+9}-3\right)^2}{\beta  (\zeta -1) (\zeta +15)^2}-6 s(r) s'(r)\right)}{6 \kappa  r^3 a(t)^4 \left(\zeta +2 \sqrt{\frac{6 \beta  (\zeta -1) (\zeta +15) \left(3 r^2 \dot{a}(t)^2+2 r s(r) s'(r)+s(r)^2\right)}{\kappa ^2 r^2 a(t)^2}+9}+9\right)}.
\end{equation}

Also, the scale factor in (\ref{41}), i.e. $ a(t) = \left( a_0 t - a_1 \right)^{2/3} $, where $ a_0$ and $a_1$ are constants, is real and defined only when $a_0 t - a_1 \geq 0$, which implies the domain $t \geq \frac{a_1}{a_0}$ for $a_0 > 0$, and $t \leq \frac{a_1}{a_0}$ for $a_0 < 0$. Hence, $\dot{a}(t) = \frac{2}{3} a_0 \left( a_0 t - a_1 \right)^{-1/3}$, indicates that the scale factor is monotonically increasing for $a_0 > 0$ and decreasing for $a_0 < 0$. Therefore, the model describes an expanding wormhole for $a_0 > 0$ and a contracting wormhole for $a_0 < 0$. In both cases, the scale factor vanishes at $t = \frac{a_1}{a_0}$, corresponding to a cosmological singularity.

\begin{figure*}[t!]
\begin{minipage}{\textwidth}
\includegraphics[width=\textwidth]{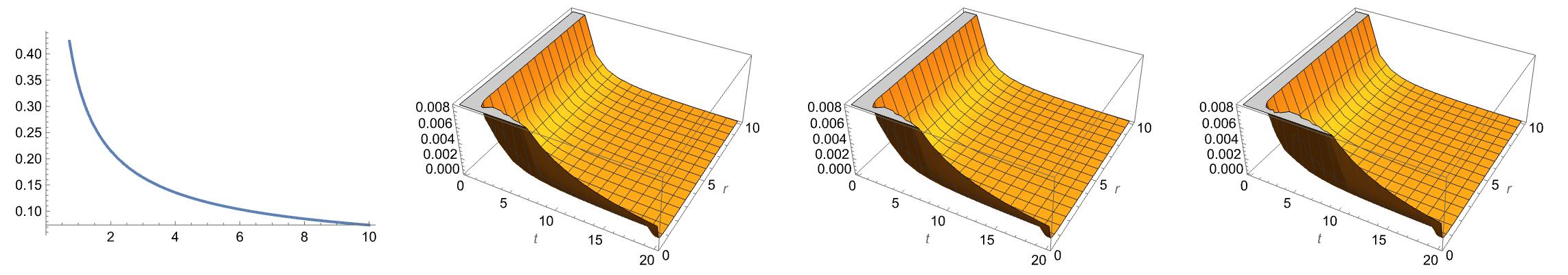}
\caption{\label{fig-s1} The behavior of $\rho$, $\rho+ P_r$ , and $\rho+ P_l$, from left to right, is illustrated for $0.2 < r < 10$ and $0 < t < 20 $, corresponding to Eqs.\eqref{74} ,\eqref{75}, and \eqref{76}, where $s(r)=\left(\frac{r}{r_0} \right)^{1/\alpha}$.
The arbitrary constants are considered as $\beta=2$, $\alpha=-3$, $a_0=2$, $a_1=-2$, $\zeta=2$, $r_0=0.2$, and $\kappa=8 \pi$, in accordance with the constraints given in \eqref{weccas200-02}.}
\end{minipage}
\end{figure*}

\begin{figure*}[t!]
\begin{minipage}{\textwidth}
\includegraphics[width=\textwidth]{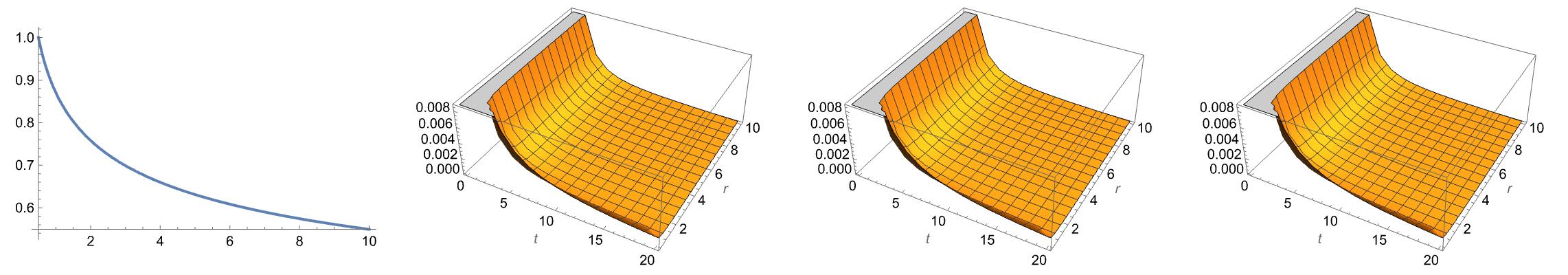}
\caption{\label{fig-s2} The behavior of $B(r)/r$, $\rho$, $\rho+P_r$ , and $\rho+P_l$, from left to right, is indicated for $0.5 < r < 10$ and $0 < t < 20$, corresponding to Eqs.\eqref{78} ,\eqref{79}, and \eqref{80}, where $s(r)=\left(\frac{r}{r_0} \right)^{n/2}$.
The arbitrary constants are set as $\beta=2$, $n=-0.2$, $a_0=2$, $a_1=-2$, $\zeta=2$, $r_0=0.5$, and $\kappa=8 \pi$, ensuring they remain within the limits specified in  \eqref{weccas201-02}.}
\end{minipage}
\end{figure*}

Exploiting the scale factor \eqref{41}, we can characterize the behavior of matter components for the aforementioned examples of $s(r)$ as follows:
\begin{itemize}
\item[$\bullet$] \label{wec-g1} \textit{ $\mathbf{s(r)=\left(\frac{r}{r_0} \right)^{1/\alpha}}$}\\
        \\
In this case, we find
\begin{equation}
    \label{74}
       \rho=\frac{\kappa  \left(\sqrt{9+\boldsymbol{\chi}_2}-3\right)}{\beta  (\zeta -1) (\zeta +15)}, 
\end{equation}
\begin{equation}
    \label{75}
    \begin{aligned}
        \rho+P_r=\frac{(\zeta +15)}{3 \alpha  \kappa  r^2 (a_0 t-a_1)^{8/3} \left(\zeta +2 \sqrt{9+\boldsymbol{\chi}_2}+9\right)}
        &\left( 3(a_0 t-a_1) (\alpha +2) \left(\frac{r}{r_0}\right)^{2/\alpha } \sqrt[3]{a_0 t-a_1} \right.\\
        &\left.+ 4 \alpha  r^2 a_0^2 (a_0 t-a_1)^{2/3}+3 \alpha  \left(\left(\frac{r}{r_0}\right)^{2/\alpha }-1\right) \left(\frac{r}{r_0}\right)^{2/\alpha } \right),
    \end{aligned}
\end{equation}
\begin{equation}
    \label{76}
    \begin{aligned}
        \rho+P_l= \frac{(\zeta +15)}{3 \alpha  \kappa  r^4 (a_0 t-a_1)^{8/3} \left(\zeta +2 \sqrt{9+\boldsymbol{\chi}_2}+9\right)}
        &\left(4 \alpha  r^4 a_0^2 (a_0 t-a_1)^{2/3}-3 \left(\frac{r}{r_0}\right)^{2/\alpha } \right.\\
        & \left. ~~+3 (\alpha +2) r^2 \left(\frac{r}{r_0}\right)^{2/\alpha } (a_0 t-a_1)^{4/3} \right),
    \end{aligned}
\end{equation}
Where $\boldsymbol{\chi}_2$ is given by
\begin{equation}
    \label{chi4}
    \boldsymbol{\chi}_2=\frac{2 \beta  (\zeta -1) (\zeta +15) \left(4 \alpha  r^2 a_0^2+3 (\alpha +2) \left(\frac{r}{r_0}\right)^{2/\alpha } (a_0 t-a_1)^{2/3}\right)}{\alpha  \kappa ^2 r^2 (a_1-a_0 t)^2}.
\end{equation}
Here, a viable wormhole configuration that holds the WEC must fall within the following permissible range of parameters
\begin{equation}
    \label{weccas200}
    \beta <0, \, ~~ -15<\zeta <1,~~ a_0 \neq 0, ~~ a_1<0, ~~ \alpha \le-2, ~~ r_0>0,
\end{equation}
or
\begin{equation}
    \label{weccas200-02}
    \beta >0, ~~ \zeta >1,~~ a_0 \neq 0, ~~ a_1<0, ~~ \alpha \le-2, ~~ r_0>0.
\end{equation}
Accordingly, Figure \ref{fig-s1} demonstrates that both the WEC and flatness condition are satisfied for a specific set of parameters in this scenario.
\item[$\bullet$] \label{wec-g2} \textit{$\mathbf{s(r)=\left(\frac{r}{r_0} \right)^{n/2}}$}\\
        \\
In this case, we have 
\begin{equation}
    \label{78}
    \rho=\frac{\kappa  \left(\sqrt{9+\boldsymbol{\chi}_3}-3\right)}{\beta  (\zeta -1) (\zeta +15)},
\end{equation}
\begin{equation}
    \label{79}
    \begin{aligned}
        \rho+P_r=\frac{(\zeta +15)}{3 \kappa r^2 (a_0 t-a_1)^{8/3} \left(\zeta +2 \sqrt{9+\boldsymbol{\chi}_3}+9\right)}
        &\left( 3 \left(\left(\frac{r}{r_0}\right)^n-1\right) \left(\frac{r}{r_0}\right)^n+4 r^2 a_0^2 (a_0 t-a_1)^{2/3}\right.\\
        &\left.~~+3 (n+1) \left(\frac{r}{r_0}\right)^n (a_0 t-a_1)^{4/3} \right),
    \end{aligned}
\end{equation}
\begin{equation}
    \label{80}
    \begin{aligned}
        \rho+P_l=\frac{(\zeta +15)}{6  \kappa  r^4 (a_0 t-a_1)^{8/3} \left(\zeta +2 \sqrt{9+\boldsymbol{\chi}_3}+9\right)}
        &\left(6 r^2 \left(\frac{r}{r_0}\right)^n (a_0 t-a_1)^{4/3}+8 r^4 a_0^2 (a_0 t-a_1)^{2/3} \right.\\
        & \left. ~~-3 n \left(\frac{r}{r_0}\right)^n \left(1-2 r^2 (a_0 t-a_1)^{4/3}\right) \right),
    \end{aligned}
\end{equation}
where $\boldsymbol{\chi}_3$ is given by
\begin{equation}
    \label{chi5}
    \boldsymbol{\chi}_3=\frac{2 \beta  (\zeta -1) (\zeta +15) \left(3 (n+1) \left(\frac{r}{r_0}\right)^n (a_0 t-a_1)^{2/3}+4 r^2 a_0^2\right)}{\kappa ^2 r^2 (a_1-a_0 t)^2}.
\end{equation}
It is observed that wormholes satisfying the WEC are attainable within specific parameter ranges as follows:
\begin{equation}
    \label{weccas201}
    \beta <0, ~~ -15<\zeta <1,~~ r_0>0, 
    \begin{cases}
        n<1 ~~~~ &-\frac{8\sqrt{3}}{9} \, \sqrt{\frac{a_0^6 r_0^6}{(n+1)^3}}<a_1<0,~~ a_0 \neq0, \\ \\
        -1 \leq n <0, ~~~~ &a_1<0, ~~ a_0 \neq0,
    \end{cases}
\end{equation}
or
\begin{equation}
    \label{weccas201-02}
    \beta >0, ~~ \zeta >1,~~ r_0>0, 
    \begin{cases}
        n<1 ~~~~ &-\frac{8\sqrt{3}}{9} \, \sqrt{\frac{a_0^6 r_0^6}{(n+1)^3}}<a_1<0,~~ a_0, \neq0 \\ \\
        -1 \leq n <0, ~~~~ &a_1<0, ~~ a_0 \neq0,
    \end{cases}
\end{equation}
As depicted in Fig. \ref{fig-s2}, specific parameter values within the aforementioned ranges satisfy both the WEC and flatness conditions.
\\
    \item[$\bullet$] \label{wec-g3}\textit{$\mathbf{s(r)=\sqrt{\frac{1}{1+r-r_0}}}$}\\
\\
This form of $s(r)$ leaves us
\begin{equation}
    \label{82}
    \rho=\frac{\kappa  \left(\sqrt{9+\boldsymbol{\chi}_4}-3\right)}{\beta  (\zeta -1) (\zeta +15)},
\end{equation}
\begin{equation}
    \label{83}
    \begin{aligned}
        \rho+P_r=\frac{(\zeta +15)}{3 \kappa  r^2 (r-r_0+1)^2 (a_0 t-a_1)^{8/3} \left(\zeta +2 \sqrt{9+\boldsymbol{\chi}_4}+9\right) }
        &\left\{ r \left(4 r a_0^2 (r-r_0+1)^2 (a_0 t-a_1)^{2/3}-3\right) \right.\\
        & \left. ~~+3 \left(-r_0 (a_0 t-a_1)^{4/3}+r_0+(a_0 t-a_1)^{4/3}\right) \right\},
    \end{aligned}
\end{equation}
\begin{equation}
    \label{84}
    \begin{aligned}
        \rho+P_l=\frac{(\zeta +15)}{6 r^3 (r-r_0+1)^2 (a_0 t-a_1)^{8/3} \left(\zeta +2 \sqrt{9+\boldsymbol{\chi}_4}+9\right)}
        &\left\{3+ 2 r \sqrt[3]{a_0 t-a_1}\bigg( 3 (r_0-1) a_1 \right. \\
        &  \left. ~+a_0 \left(4 r^2 a_0 (r-r_0+1)^2 \sqrt[3]{a_0 t-a_1}-3 (r_0-1) t\right) \bigg) \right\},
    \end{aligned}
\end{equation}
where $\boldsymbol{\chi}_4$ is defined as
\begin{equation}
    \label{chi6}
    \boldsymbol{\chi}_4=\frac{8 \beta  (\zeta -1) (\zeta +15) a_0^2}{\kappa ^2 (a_1-a_0 t)^2}-\frac{6 \beta  (\zeta -1) (\zeta +15) (r_0-1)}{\kappa ^2 r^2 (r-r_0+1)^2 (a_0 t-a_1)^{4/3}}.
\end{equation}
The parameter restrictions necessary to satisfy the weak energy condition and flatness condition read as
\begin{equation}
    \label{weccas202}
    \beta <0,~~ -15<\zeta <1, 
    \begin{cases}
        r_0>1 ~~~~ &-\frac{8\sqrt{3}}{9} \, \sqrt{-\frac{a_0^6 r_0^6}{(r_0-1)^3}}<a_1<0,~~ a_0 \neq0, \\ \\
        0<r_0 \leq1, ~~~~ &a_1<0, ~~ a_0 \neq0,
    \end{cases}
\end{equation}
or
\begin{equation}
    \label{weccas20202}
    \beta >0, ~~ \zeta >1, 
    \begin{cases}
        r_0>1 ~~~~ &-\frac{8\sqrt{3}}{9} \, \sqrt{-\frac{a_0^6 r_0^6}{(r_0-1)^3}}<a_1<0,~~ a_0 \neq0, \\ \\
        0<r_0 \leq1, ~~~~ &a_1<0, ~~ a_0 \neq0.
    \end{cases}
\end{equation}
Figure \ref{fig-s3} shows a specific set of parameter values underscoring these constraints.
\\
    \item[$\bullet$] \label{wec-g4} 
    \textit{$\mathbf{s(r)=\sqrt{\epsilon (1-\frac{r_0}{r})+\frac{r_0}{r}}}$}\\
\\
Assuming $s(r)$ with this form, yields matter components as follows
\begin{equation}
    \label{86}
    \rho=\frac{\kappa  \left(\sqrt{9+\boldsymbol{\chi}_5}-3\right)}{\beta  (\zeta -1) (\zeta +15)}
\end{equation}
\begin{equation}
    \label{87}
    \begin{aligned}
\rho+P_r=\frac{(\zeta+15)\left(4 r^4 a_0^2 (a_0 t-a_1)^{2/3}+3 r_0^2 (\epsilon -1)^2+3 r^2 \epsilon  \left((a_0 t-a_1)^{4/3}+\epsilon -1\right)-3 r r_0 (\epsilon -1) (2 \epsilon -1) \right)}{3 \kappa r^4 (a_0 t-a_1)^{8/3} \left( 9+ \zeta +2 \sqrt{9+\boldsymbol{\chi}_5}\right)}
    \end{aligned}
\end{equation}
\begin{equation}
    \label{88}
    \begin{aligned}
\rho+P_l=\frac{(\zeta+15)\left(8 r^5 a_0^2 (a_0 t-a_1)^{2/3}+6 r^3 \epsilon  (a_0 t-a_1)^{4/3}-3 r_0 \epsilon +3 r_0 \right)}{6 \kappa r^5 (a_0 t-a_1)^{8/3} \left( 9+ \zeta +2 \sqrt{9+\boldsymbol{\chi}_5}\right)}
    \end{aligned}
\end{equation}
where $\boldsymbol{\chi}_5$ symbolizes following expression
\begin{equation}
    \label{chi7}
\boldsymbol{\chi}_5=\frac{2 \beta  (\zeta -1) (\zeta +15) \left(4 r^2 a_0^2+3 \epsilon  (a_0 t-a_1)^{2/3}\right)}{\kappa ^2 r^2 (a_1-a_0 t)^2}.
\end{equation}
The corresponding parameter constraints for validity of WEC and flatness condition for this form of $s(r)$ are
\begin{equation}
    \label{weccas203}
    \beta <0, ~~ -15<\zeta <1,~~ r_0>0, 
    \begin{cases}
        \epsilon<0 ~~~~ &-\frac{8\sqrt{3}}{9} \, \sqrt{-\frac{a_0^6 r_0^6}{\epsilon^3}}<a_1<0,~~ a_0 \neq0, \\ \\
        0\leq \epsilon \leq 1, ~~~~ &a_1<0, ~~ a_0 \neq0,
    \end{cases}
\end{equation}
or
\begin{equation}
    \label{weccas20302}
    \beta >0, ~~ \zeta >1,~~ r_0>0, 
    \begin{cases}
        \epsilon<0 ~~~~ &-\frac{8\sqrt{3}}{9} \, \sqrt{-\frac{a_0^6 r_0^6}{\epsilon^3}}<a_1<0,~~ a_0 \neq0, \\ \\
        0\leq \epsilon \leq 1, ~~~~ &a_1<0, ~~ a_0 \neq0.
    \end{cases}
\end{equation}
Additionally, Figure \ref{fig-s4} demonstrates the adherence to the WEC for parameters within these feasible ranges.
In the late-time limit  ($t \to \infty$), the exact solutions we obtained indicate that the energy density $ \rho$ and the pressure components $P_r $ and $ P_l $ asymptotically vanish for $a_0>0$. This behavior is physically consistent, as the corresponding exact solution for the scale factor exhibits growth with time, reflecting an expanding wormhole spacetime for $a_0>0$. The behavior of $\rho$, $P_r$, and $P_l$ as $t \to \infty$ is shown in Fig \ref{re-fig2infty}.
\end{itemize}

\begin{figure*}[t!]
\begin{minipage}{\textwidth}
\includegraphics[width=\textwidth]{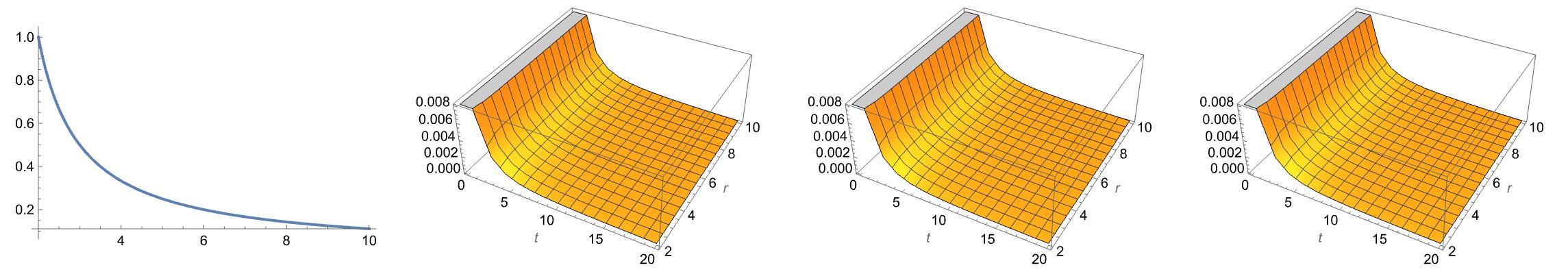}
\caption{\label{fig-s3} The behavior of $B(r)/r$, $\rho$, $\rho+P_r$ , and $\rho+P_l$, from left to right, is shown for $2 < r < 10$ and $0 < t < 20$, corresponding to Eqs.\eqref{82} ,\eqref{83}, and \eqref{84}, where $s(r)=\sqrt{\frac{1}{1+r-r_0}}$.
The arbitrary constants are chosen as $\beta=2$, $\zeta=2$, $a_0=2$, $a_1=-2$, $r_0=2$, and $\kappa=8 \pi$, in alignment with the constraints of \eqref{weccas20202}.}
\end{minipage}
\end{figure*}
\begin{figure*}[t!]
\begin{minipage}{\textwidth}
\includegraphics[width=\textwidth]{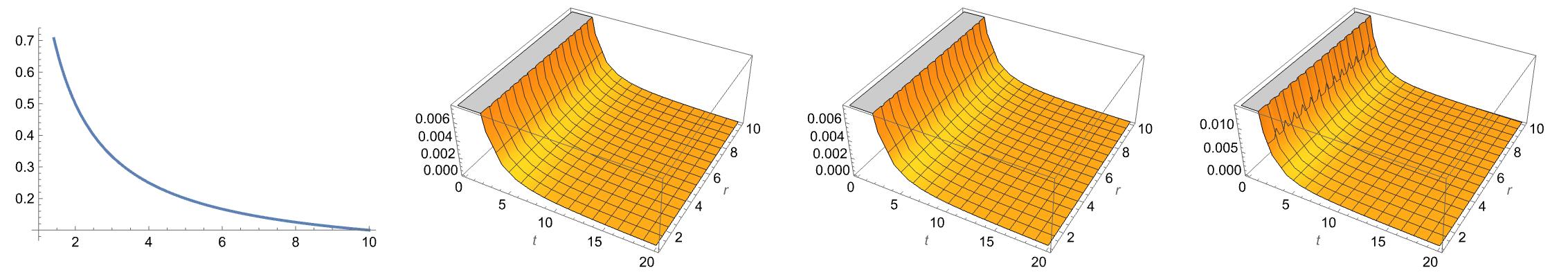}
\caption{\label{fig-s4} The behavior of $B(r)/r$, $\rho$, $\rho+P_r$ , and $\rho+P_l$, from left to right, is indicated for $1 < r < 10$ and $0 < t < 20$, corresponding to Eqs.\eqref{86} ,\eqref{87}, and \eqref{88}, where $s(r)=\sqrt{\epsilon (1-\frac{r_0}{r})+\frac{r_0}{r}}$.
The arbitrary constants are taken as $\beta=2$, $\epsilon=-2$, $a_0=2$, $a_1=-0.3$, $\zeta=2$, $r_0=1$, and $\kappa=8 \pi$, consistent with the conditions established in \eqref{weccas20302}.}
\end{minipage}
\end{figure*}
\begin{figure*}[h!]
\begin{minipage}{\textwidth}
\includegraphics[width=0.75\textwidth]{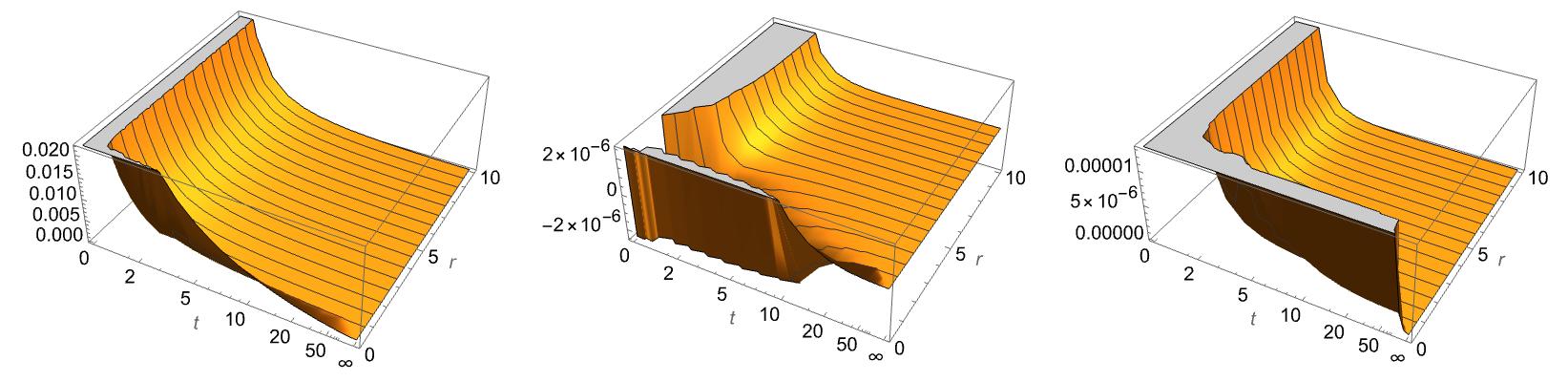}
\caption{\label{re-fig2infty} The behavior of $\rho$, $P_r$ , and $P_l$, from left to right, is indicated for $0.2 < r < 10$ and $0 < t < \infty $, corresponding to Eqs.\eqref{rhocase265} ,\eqref{prcase266}, and \eqref{Plcase267}, exploiting the scale factor \eqref{41} in the case of $s(r)=\left(\frac{r}{r_0} \right)^{1/\alpha}$ \eqref{37}.
The arbitrary constants are considered as $\beta=2$, $\alpha=-3$, $a_0=2$, $a_1=-2$, $\zeta=2$, $r_0=0.2$, and $\kappa=8 \pi$, in accordance with the constraints given in \eqref{weccas200-02} and same as Fig \ref{fig-s1}.}
\end{minipage}
\end{figure*}
As we mentioned earlier, for all $s(r)$ cases studied, $T_{\mu \nu}^{eff}$ violates the WEC within determined parameter ranges. However, as shown in Figures \ref{fig-s1}-\ref{fig-s4}, the ordinary energy-momentum tensor $T_{\mu \nu}$ satisfies the WEC — unlike in GR, where such a violation would necessarily occur.

As the final remark, in the present study, we have shown that traversable wormholes in $f(R,T)$ gravity can be supported by phantom energy ($\omega < -1$). In \cite{stablewh}, Lobo investigated the stability of such phantom wormholes—an issue of fundamental importance—by analyzing linearized perturbations around static solutions and incorporating the momentum flux term in the conservation identity. Their results demonstrated that stable equilibrium configurations can exist under specific conditions. Moreover, it was shown that stability can be enhanced by appropriately tuning the wormhole throat radius and other parameters. Given the broader range of wormhole configurations identified in the present work, we plan to investigate the stability of these solutions in future studies.
\section{\label{Conclusion} Conclusion}
This study has explored dynamical wormhole solutions within the framework of $f(R,T) = R + \beta T^2$ gravity, advancing our understanding of traversable wormholes in modified gravity theories. By introducing a quadratic dependence on the trace of the stress-energy tensor, we have demonstrated that such wormholes can be sustained without violating the WEC for ordinary matter—a striking departure from General Relativity GR, where exotic matter is typically required. 
\\
The $T^2$ term introduces high nonlinearities into the field equations, making them analytically intractable with standard methods. We overcome this challenge by developing a novel solution procedure and incorporating two general EoS that handles these nonlinearities, providing exact analytical solutions for the $f(R, T)$ theory.
By reformulating the field equations in terms of an effective energy-momentum tensor, we bypass the need for ad hoc ansatzes (e.g., predefined shape functions) that were considered in previous static wormhole studies considering $T^2$ term \cite{3-1, 3-2, intro46, intrort61, 7-2}. This allows us to derive exact analytical solutions for both the scale factor $a(t)$ and the shape function $B(r)$.
The framework is general enough to accommodate arbitrary equations of state (EoS). We consider two distinct EoS to ensure robustness; Case I: a general EoS linking $\rho$, $P_r$, and Case II: $P_l$, and (ii) a simpler EoS relating only $P_r$ and $P_l$.
This flexibility ensures our results are not restricted to a very specific EoS.
The developed methodology is not limited to the investigated particular case of $f(R, T)$ theory but extends to other modified gravity theories with higher-order matter couplings. This paves the way for finding other novel exact solutions in these theories.
\\
Integrating the corresponding field equations, we derive a general class of solutions for the scale factor and shape function of the wormhole. We examine the obtained solutions versus the conditions under which both the WEC and asymptotic flatness condition are satisfied in the context of this special case of $f(R,T)$ gravity. The consideration of two distinct EoS provides insights into the size and energy violation characteristics of the corresponding solutions. In particular, utilizing the EoS incorporating only $P_r$ and $P_l$ leaves the wormhole throat independent of any parameter and hence leads to a wormhole of arbitrary size. In other words, energy conditions are valid for a wide range of values of $r$, in line with the result that Moraes and Sahoo concluded in \cite{intro46} for a static wormhole. On the other hand, adopting the EoS with $\rho,
P_r$ , and $P_l$ impose constraints on the size of the wormhole throat through the EoS parameters $\omega$ and $\gamma$. We present a comprehensive classification of solutions and their corresponding constraints, which ensure compliance with the WEC. These are organized according to the theory’s coupling constant $\beta$ and the derived metric functions, as detailed in Table~\eqref{tab3}.
\\
It is worth mentioning that the dynamical nature of our solutions arises from the explicit time dependence of the metric, which embeds the wormhole within a cosmological context. This dynamical behavior is further enriched by the non-conservation of the ordinary energy-momentum tensor, $\nabla_\mu T^{\mu\nu} \neq 0$, a hallmark of $f(R,T)$ gravity. The additional terms in $T^{eff}_{\mu\nu}$—interpretable as imperfect fluids, quantum corrections, or an effective cosmological constant \cite{hark, conservation2, conservation3}— represent an interplay between geometry and matter source beyond GR and act as a dynamical source for the wormhole, enabling it to evade the weak energy condition violations of GR. Remarkably, while $T^{eff}_{\mu\nu}$ violates the WEC, the ordinary energy-momentum tensor $T_{\mu\nu}$ can fully satisfy it. The solutions obtained for the ordinary energy-momentum tensor components in both cases I and II (Eqs. (\eqref{rho3.41}- \eqref{pl3.43} and \eqref{rhocase265}-\eqref{Plcase267}) follow the equations in \eqref{10} and \eqref{11}. This
underscore how $f(R,T)$ gravity redefines the matter-geometry interplay to support traversable wormholes. 
\\
For Case I, having the solution \eqref{30}, there are two classes of solutions: (i)
 If $\frac{2\omega}{3(\omega + 1)} > 0$ and $a_0 > 0$, then $\dot{a}(t) > 0$ representing an expanding wormhole,~and (ii)
 If $\frac{2\omega}{3(\omega + 1)} < 0$ and $a_0 > 0$, then $\dot{a}(t) < 0$ indicating a contracting wormhole. Hence, if $a_0>0$, for typical matter sources such as dust ($\omega = 0$) or radiation ($\omega = 1/3$), the wormhole expands as a power law in time. If $\omega < -1$, corresponding to phantom energy, the scale factor exhibits super-accelerated expansion. Also, when $ -1<\omega<0$, the WEC is not satisfied for the solutions obtained, and hence this range should be excluded. For Case II, having solution (\ref{41}),  the model describes an expanding wormhole for $a_0 > 0$ and a contracting wormhole for $a_0 < 0$. In both cases I and II, the scale factor vanishes at $t =\mp \frac{a_1}{a_0}$, respectively, corresponding to a cosmological singularity.
\\ 
In the late-time limit  ($t \to \infty$), the solutions respecting the WEC in both cases I and II indicate that the energy density $ \rho$ and the pressure components $P_r $ and $ P_l $ asymptotically vanish for $a_0>0$. This behavior is physically consistent, as the corresponding exact solution for the scale factor exhibits growth with time, reflecting an expanding wormhole spacetime for $a_0>0$. 
\\
Our findings also highlight the broader applicability of the techniques developed here. The solution procedure for handling the nonlinearities introduced by the $T^2$ term can be extended to other modified gravity theories with higher-order matter couplings. This paves the way for future investigations in finding new solutions, such as black holes, and the astrophysical implications of such solutions in various modified theories. The framework is also general enough to accommodate arbitrary EoS for the matter content in spacetime.
\renewcommand{\arraystretch}{3.1}
\begin{table*}[ht!]
\centering
\resizebox{\textwidth}{!}{\begin{tabular}{ |c|c|c|c|c|c|c|c|}
\hline
  metric functions &$\beta$ & $\zeta$ &$\omega$ & $\gamma$ &$r_0$ & $\alpha, n, \epsilon$ & $a_0$,$a_1$ \\
\hline
\multirow{8}{4cm}{\centering \makecell{$B(r)= r_0 \left(\frac{r}{r_0}\right)^{\frac{(\gamma -1) \omega}{1+\gamma ( \omega+2)}}$,\\ $a(t)= (a_0 t+ a_1)^{\frac{2 \omega }{3 (\omega +1)}}$}}& \multirow{4}{*}{$\beta<0$}& \multirow{4}{*}{$-15< \zeta <1$}& $\omega<-3$& $\frac{\omega -1}{2 \omega +2}<\gamma <\frac{1}{2} (-\omega -1)$& $r_0 > \frac{\sqrt{3}}{2} \sqrt{\frac{(\gamma -1) a_1^2 (\omega +1)^2 a_1^{-\frac{4 \omega }{3 (\omega +1)}}}{a_0^2 (\gamma  (\omega +2) \omega +\omega )}}$&-&\multirow{3}{*}{\makecell{$a_0>0$,\\ $a_1>0$}}\\
\cline{4-7}
 & & & $-3<\omega<-2$ &$\frac{1}{2} (-\omega -1)<\gamma <\frac{\omega -1}{2 \omega +2}$ & $r_0>\frac{\sqrt{3}}{2} | \omega +1|  \sqrt{\frac{(\gamma -1) a_1^2 a_1^{-\frac{4 \omega }{3 (\omega +1)}}}{a_0^2 (\gamma  (\omega +2) \omega +\omega )}}$&-& \\
 \cline{4-7}
 & & & $-2 \leq \omega <-1$ & $\frac{1}{2} (-\omega -1)<\gamma <\frac{\omega -1}{2 \omega +2}$ &$r_0 \geq \frac{\sqrt{6}}{4} \sqrt{-\frac{(\omega +1)^2 (2 \gamma +\omega +1) a_1^{-\frac{4 \omega }{3 (\omega +1)}}}{a_0^2 \omega  (\gamma  (\omega +2)+1) a_1^{-\frac{2 (5 \omega +3)}{3 (\omega +1)}}}}$&-& \\
 \cline{4-8}
 & & & $\omega>0$ &$\frac{\omega -1}{2 \omega +2}< \gamma \leq 1$&$r_0>\frac{1}{2} \sqrt{3} \sqrt{\frac{(\gamma -1) a_1^2 (\omega +1)^2 a_1^{-\frac{4 \omega }{3 (\omega +1)}}}{a_0^2 (\gamma  (\omega +2) \omega +\omega )}}$&-& \makecell{$a_0>0$,\\ $a_1>0$}\\
 \cline{2-8}
 & \multirow{4}{*}{$\beta>0$}& \multirow{4}{*}{$\zeta >1$}& $\omega<-3$& $\frac{\omega -1}{2 \omega +2}<\gamma <\frac{1}{2} (-\omega -1)$& $r_0 > \frac{\sqrt{3}}{2} \sqrt{\frac{(\gamma -1) a_1^2 (\omega +1)^2 a_1^{-\frac{4 \omega }{3 (\omega +1)}}}{a_0^2 (\gamma  (\omega +2) \omega +\omega )}}$&-&\multirow{3}{*}{\makecell{$a_0>0$,\\ $a_1>0$}}\\
\cline{4-7}
 & & & $-3<\omega<-2$ &$\frac{1}{2} (-\omega -1)<\gamma <\frac{\omega -1}{2 \omega +2}$ & $r_0>\frac{\sqrt{3}}{2} | \omega +1|  \sqrt{\frac{(\gamma -1) a_1^2 a_1^{-\frac{4 \omega }{3 (\omega +1)}}}{a_0^2 (\gamma  (\omega +2) \omega +\omega )}}$&-& \\
 \cline{4-7}
 & & & $-2 \leq \omega <-1$ & $\frac{1}{2} (-\omega -1)<\gamma <\frac{\omega -1}{2 \omega +2}$ &$r_0 \geq \frac{\sqrt{6}}{4} \sqrt{-\frac{(\omega +1)^2 (2 \gamma +\omega +1) a_1^{-\frac{4 \omega }{3 (\omega +1)}}}{a_0^2 \omega  (\gamma  (\omega +2)+1) a_1^{-\frac{2 (5 \omega +3)}{3 (\omega +1)}}}}$&-& \\
 \cline{4-8}
 & & & $\omega>0$ &$\frac{\omega -1}{2 \omega +2}< \gamma \leq 1$&$r_0>\frac{1}{2} \sqrt{3} \sqrt{\frac{(\gamma -1) a_1^2 (\omega +1)^2 a_1^{-\frac{4 \omega }{3 (\omega +1)}}}{a_0^2 (\gamma  (\omega +2) \omega +\omega )}}$&-& \makecell{$a_0>0$,\\ $a_1>0$}\\
 \hline
\multirow{2}{3cm}{\centering \makecell{$s(r)=\left(\frac{r}{r_0} \right)^{1/\alpha}$,\\$a(t)=(a_0 \, t - a_1)^{2/3}$}}& $\beta<0$ & $-15<\zeta <1$ &\multirow{2}{*}{$-$}&\multirow{2}{*}{$-$} & \multirow{2}{1cm}{\centering$r_0>0$} & \multirow{2}{*}{ $\alpha \leq -2$} &\multirow{2}{1cm}{\centering \makecell{$a_0\ne 0$,\\ $a_1<0$}}\\
\cline{2-3}
  & $\beta>0$ & $\zeta>1$& & & & & \\
\hline
\multirow{4}{3cm}{\centering \makecell{$s(r)=\left(\frac{r}{r_0}\right)^{n/2}$,\\$a(t)=(a_0 \, t - a_1)^{2/3}$}}& \multirow{2}{1cm}{\centering $\beta<0$ }&  \multirow{2}{*}{\centering$-15<\zeta <1$}& \multirow{2}{*}{$-$} & \multirow{2}{*}{$-$} & \multirow{2}{*}{\centering $r_0>0$}& $n<-1$ &  \makecell{$a_0\ne 0$,\\ $-\frac{8\sqrt{3}}{9} \, \sqrt{-\frac{a_0^6 r_0^6}{(n+1)^3}}<a_1<0$}\\ 
\cline{7-8}
 & & & & & & $-1 \leq n<0$ &\makecell{$a_0\ne 0$,\\$a_1<0$}\\
\cline{2-8}
  & \multirow{2}{1cm}{\centering $\beta>0$ }& \multirow{2}{*}{\centering$\zeta>1$}& \multirow{2}{*}{$-$} & \multirow{2}{*}{$-$} & \multirow{2}{1cm}{\centering $r_0>0$}& $n<-1$ & \makecell{$a_0\ne 0$,\\ $-\frac{8\sqrt{3}}{9} \, \sqrt{-\frac{a_0^6 r_0^6}{(n+1)^3}}<a_1<0$}\\
\cline{7-8}
 & & & & & & $-1 \leq n <0$ & \makecell{$a_0\ne 0$,\\$a_1<0$}\\
\hline
\multirow{4}{3cm}{\centering \makecell{ $s(r)=\sqrt{\frac{1}{r-r_0+1}}$,\\$a(t)=(a_0 \, t - a_1)^{2/3}$}}& \multirow{2}{1cm}{\centering $\beta<0$ }&  \multirow{2}{*}{\centering$-15<\zeta<1$} &\multirow{2}{*}{$-$}&\multirow{2}{*}{$-$}&  $r_0>1$& $-$ &\makecell{$a_0\ne 0$,\\ $-\frac{8\sqrt{3}}{9} \, \sqrt{\frac{a_0^6 r_0^6}{(r_0-1)^3}}<a_1<0$}\\
\cline{6-8}
 & & & & & $0<r_0\leq 1$ & $-$& \makecell{$a_0\ne 0$,\\ $a_1<0$}\\
\cline{2-8}
  & \multirow{2}{1cm}{\centering $\beta>0$ } & \multirow{2}{*}{\centering$\zeta>1$}& \multirow{2}{*}{$-$}& \multirow{2}{*}{$-$}& $r_0>1$& $-$ & \makecell{$a_0\ne 0$,\\ $-\frac{8\sqrt{3}}{9} \, \sqrt{\frac{a_0^6 r_0^6}{(r_0-1)^3}}<a_1<0$}\\
\cline{6-8}
 & & & & & $0<r_0\leq 1$ & $-$ & \makecell{$a_0\ne 0$,\\ $a_1<0$}\\
\hline
\multirow{4}{3cm}{\centering \makecell{$s(r)=\sqrt{\epsilon  \left(1-\frac{r_0}{r}\right)+\frac{r_0}{r}}$,\\$a(t)=(a_0 \, t - a_1)^{2/3}$}}& \multirow{2}{1cm}{\centering $\beta<0$ }&  \multirow{2}{*}{\centering$-15<\zeta<1$} & \multirow{2}{*}{$-$}& \multirow{2}{*}{$-$}& \multirow{2}{*}{\centering $r_0>0$}& $\epsilon<0$ & \makecell{$a_0\ne 0$,\\ $-\frac{8\sqrt{3}}{9} \, \sqrt{-\frac{a_0^6 r_0^6}{\epsilon^3}}<a_1<0$}\\
\cline{7-8}
 & & & & & & $0\leq \epsilon \leq 1$ & \makecell{$a_0\ne 0$,\\ $a_1<0$}\\
\cline{2-8}
  & \multirow{2}{1cm}{\centering $\beta>0$ } & \multirow{2}{*}{\centering$\zeta>1$}& \multirow{2}{*}{$-$}& \multirow{2}{*}{$-$}& \multirow{2}{*}{\centering $r_0>0$}&  $\epsilon<0$ & \makecell{$a_0\ne 0$,\\ $-\frac{8\sqrt{3}}{9} \, \sqrt{-\frac{a_0^6 r_0^6}{\epsilon^3}}<a_1<0$}\\
\cline{7-8}
 & & & & & & $0\leq \epsilon \leq 1$ & \makecell{$a_0\ne 0$,\\ $a_1<0$}\\
\hline
\end{tabular}}
\caption{The list of obtained solutions, along with the constraints ensuring compliance with the WEC.}
\label{tab3}
\end{table*}
\newpage
\section{\label{references}References}
\newcommand{\bibTitle}[1]{``#1''}
\begingroup
\let\itshape\upshape
\bibliographystyle{unsrt}

\end{document}